\documentclass{article}
\usepackage{arxiv}

\usepackage[utf8]{inputenc} 
\usepackage[T1]{fontenc}    
\usepackage{hyperref}       
\usepackage{url}            
\usepackage{booktabs}       
\usepackage{amsfonts}       
\usepackage{nicefrac}       
\usepackage{microtype}      
\usepackage{lipsum}		
\usepackage{graphicx}
\usepackage{cite}
\usepackage{multirow}

\usepackage{xcolor}

\usepackage{amsmath}

\begin{document}
%
\title{Analyzing Recursiveness in Multimodal Generative Artificial Intelligence: Stability or Divergence?}

\author{
 Javier Conde \\
 ETSI de Telecomunicación \\
 Universidad Polit\'ecnica de Madrid \\
 28040 Madrid, Spain \\
 \texttt{javier.conde.diaz@upm.es} \\
 \And
 Tobias Cheung \\
 School of Informatics \\
 University of Edinburgh \\
 EH8 9AB Edinburgh, United Kingdom \\
 \texttt{s2460940@ed.ac.uk} \\
  \And
 Gonzalo Mart\'inez \\
 Dept. Ing. Telem\'atica \\
 Universidad Carlos III de Madrid \\
 28911 Madrid, Spain \\
 \texttt{gonzmart@pa.uc3m.es} \\
 \And
 Pedro Reviriego \\
 ETSI de Telecomunicación \\
 Universidad Polit\'ecnica de Madrid \\
 28040 Madrid, Spain \\
 \texttt{pedro.reviriego@upm.es} \\
 \And
 Rik Sarkar \\
 School of Informatics \\
 University of Edinburgh \\
 EH8 9AB Edinburgh, United Kingdom \\
 \texttt{rsarkar@inf.ed.ac.uk} \\
}


\maketitle

\begin{abstract}

One of the latest trends in generative Artificial Intelligence (AI) is tools that generate and analyze content in different modalities, such as text and images, and convert information from one to the other. From a conceptual point of view, it is interesting to study whether these modality changes incur information loss and to what extent. This is analogous to variants of the classical game  telephone\footnote{\url{https://en.wikipedia.org/wiki/Chinese_whispers\#variants}}, where players alternate between describing images and creating drawings based on those descriptions leading to unexpected transformations of the original content. In the case of AI, modality changes can be applied recursively, starting from an image to extract a text that describes it; using the text to generate a second image, extracting a text that describes it, and so on. As this process is applied recursively, AI tools are generating content from one mode to use them to create content in another mode and so on. Ideally, the embeddings of all of them would remain close to those of the original content so that only small variations are observed in the generated content versus the original one. However, it may also be the case the distance to the original embeddings increases in each iteration leading to a divergence in the process and to content that is barely related to the original one. In this paper, we present the results of an empirical study on the impact of recursive modality changes using GPT-4o, a state-of-the-art AI multimodal tool, and DALL-E 3. The results show that the multimodality loop diverges from the initial image without converging to anything specific. We have observed differences depending on the type of initial image and the configuration of the models. These findings are particularly relevant due to the increasing use of these tools for content generation, reconstruction, and adaptation, and their potential implications for the content on the Internet of the future.

\end{abstract}


%

\section{Introduction}

The rise of generative AI has been marked by the emergence of chatbots based on Large Language Models (LLMs) trained to answer questions (text-to-text models) \cite{ChatGPTOverview}. In the field of image generation, models that generate images from descriptions such as DALL-E \cite{DALLE} or Stable Diffusion \cite{SD1} (known as text-to-image models) have stood out. However, new models have emerged that focus on other tasks such as image classification, Optical Character Recognition (OCR), captioning (image-to-text models), image modification, extension, or inpainting (image\&text-to-image models) \cite{inpainting_survey}, and more recently, multimodal models capable of answering questions about images (image\&text-to-text models) \cite{MultiModalSurvey}. Given this diversity of models and the increasing use of generative AI on the Internet, it is necessary to study the new interactions of the AI models and how they can impact future content on the Internet \cite{martinez2023towards}. For example, a text-to-text model can be used to summarize news from various online newspapers, with this summary an image of the news can be generated (text-to-image), and from this image, a new summary can be generated (image-to-text). Due to the lack of explainability and the variability of generative AI models, it is very difficult to determine the possible outcomes and the impact they may have on the loss, generation, or exaggeration of information. Therefore, it is necessary to define mechanisms that allow for the evaluation of the quality of the content generated by AI and the measurement of the biases, differences, and particularities that occur when introducing AI in different phases and for different tasks in content generation.

Until now, the evaluation of AI has focused on the comparison of various models and research on AI-generated content detectors \cite{LLMeval}. However, the growing use of AI tools is causing a drastic change in the generation and consumption of content on the Internet. More and more content being uploaded to the Internet is generated by or assisted by AI. For example, in scientific publications, an increase in the use of certain words like ``delve'' has been observed since the appearance of ChatGPT 3.5 \cite{kobak2024delving}. Additionally, content consumption is being altered by this technology with the emergence of search engines like Perplexity, based on Retrieval Augmented Generation (RAG) \cite{RAG} architectures that filter web page information through LLMs.
Various studies suggest that the impact of generative AI will not be limited to generation and consumption but will have a recursive effect as new AIs are trained with content generated by other AI models. Most of these works have observed a degeneration and eventually a collapse \cite{Collapse1},\cite{Collapse2},\cite{Collapse3},\cite{Collapse4}. The potential impact of this loop can be understood with a simple example, a recent study has shown that most open LLMs do not know a significant fraction of the Spanish words \cite{conde2024open}. This means that most likely they will not use those words that as a result will be less and less frequent in future training datasets and thus harder to learn by newer LLMs so they may eventually disappear from current language use.

There are other content loops created by AI tools that do not involve training newer models, such as the one explained earlier regarding the news. Another example is the recursive inpainting of images with AI tools that have been recently studied \cite{conde2024stable}. The results show that depending on the image type, the size of the fragments removed, and the number of iterations, the image obtained at the end of the process may be completely different from the original one. The development of multimodal AI tools creates loops that involve several representations of the information, for example, text and images and transformations between them that intuitively can lead to a larger loss of information. Compared to the training loop, these inference loops have received little attention and to the best of our knowledge, there is no study on the impact of recursive modality changes.

In this article, we present an empirical study of the impact of recursive modality changes on the original content using GPT-4o a state-of-the-art multimodal AI model to extract descriptions from images, and DALL-E3 to generate images from text. The main contributions of the paper are:

\begin{enumerate}
    \item To formulate the problem of recursiveness in modality changes for generative AI models,
    \item To evaluate recursive modality changes using state-of-the-art multimodal AI models.
    \item To show that the information loss due to modality changes is qualitatively different from that observed in humans and leads to degeneration as observed in AI training loops.  
\end{enumerate}

The rest of the paper is organized as follows, section \ref{Preliminaries} covers the preliminaries introducing multimodal LLMs and a discussion on the loss of information inherent to modality changes. Section \ref{Methods} describes the evaluation methodology including the models, datasets, metrics, and experiments. The results are presented and analyzed in section \ref{Results} and discussed in section \ref{Discussion}. The paper ends with the conclusions and some ideas for future work in section \ref{Conclusion}.

\section{Preliminaries}
\label{Preliminaries}

This section provides an overview of multimodal models based on LLMs as well as a discussion on the information loss caused by modality changes. 

\subsection{Multimodal AI models}

The revolution of LLMs and computer vision AI models has quickly evolved to more general large models that can process both image and text, but also video or sound and are commonly known as Multimodal Large Language Models (MLLM) 
\cite{MLLM_survey}. These models can for example answer questions on an image, convert text to video, or caption images. At a high level, an MLLM has three main elements: a modality encoder and connector, an LLM, and a modality generator. The modality encoder and connector extract the information from the modality and put it in a format compatible with the LLM. The LLM processes the information and then produces the result converted to the relevant modality by the generator \cite{MLLM_survey2}.

There are many MLLMs, both proprietary and open. Probably the most well-known is GPT4v \cite{gpt4} by OpenAI, but there are also models like Gemini from Google \cite{team2023gemini} or Claude from Anthropic. In addition to those proprietary models, there are also open models for which the code and parameters are publicly available. For example, Qwen-VL by Alibaba \cite{bai2023qwen}, InternVL \cite{InternVL}, LLaVA \cite{LLava}, or the recently introduced Llama3.1 models by Meta \cite{dubey2024llama}. As of today proprietary models outperform open models and GPT4v is probably the most widely used model. 

In many cases, as with GPT4v, the MLLM does not integrate an image generator and an independent text-to-image generator can be used to produce images from the output text. In the case of GPT4, it seems that DALL-E3 is the natural choice of image generator as it has also been developed by OpenAI, but many others are available such as Stable Difussion \cite{SD1} or Midjourney.  

In this work, since we are interested in evaluating the impact of modality changes in the content, it makes sense to study a widely used model, and at the same time one of the best-performing ones so that the results reflect the impact of changes in state-of-the-art models. These requirements make GPT4v and DALL-E3 ideal candidates for our empirical evaluation. 

\subsection{Modality changes and information loss}

\begin{figure}[h]
  \centering
  \begin{minipage}{0.45\textwidth}
    \centering
    \includegraphics[scale=1.5]{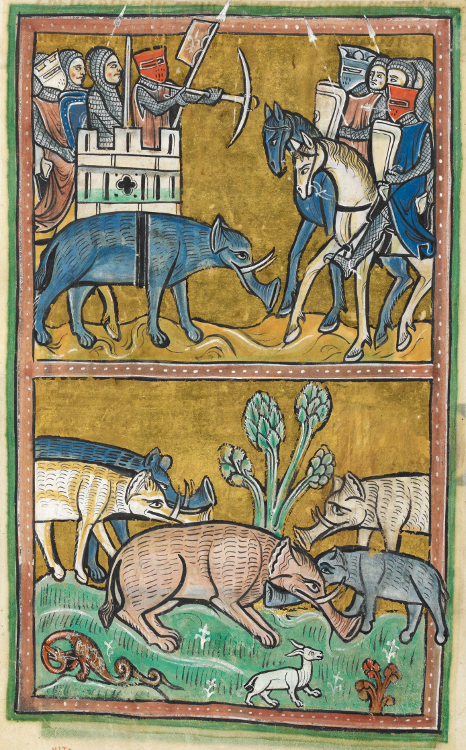}
    \caption{Illustrations of Elephants from the Rochester Bestiary.}
    \label{fig:Bestiary}
  \end{minipage}\hfill
  \begin{minipage}{0.45\textwidth}
    \centering
    \includegraphics[scale=0.8]{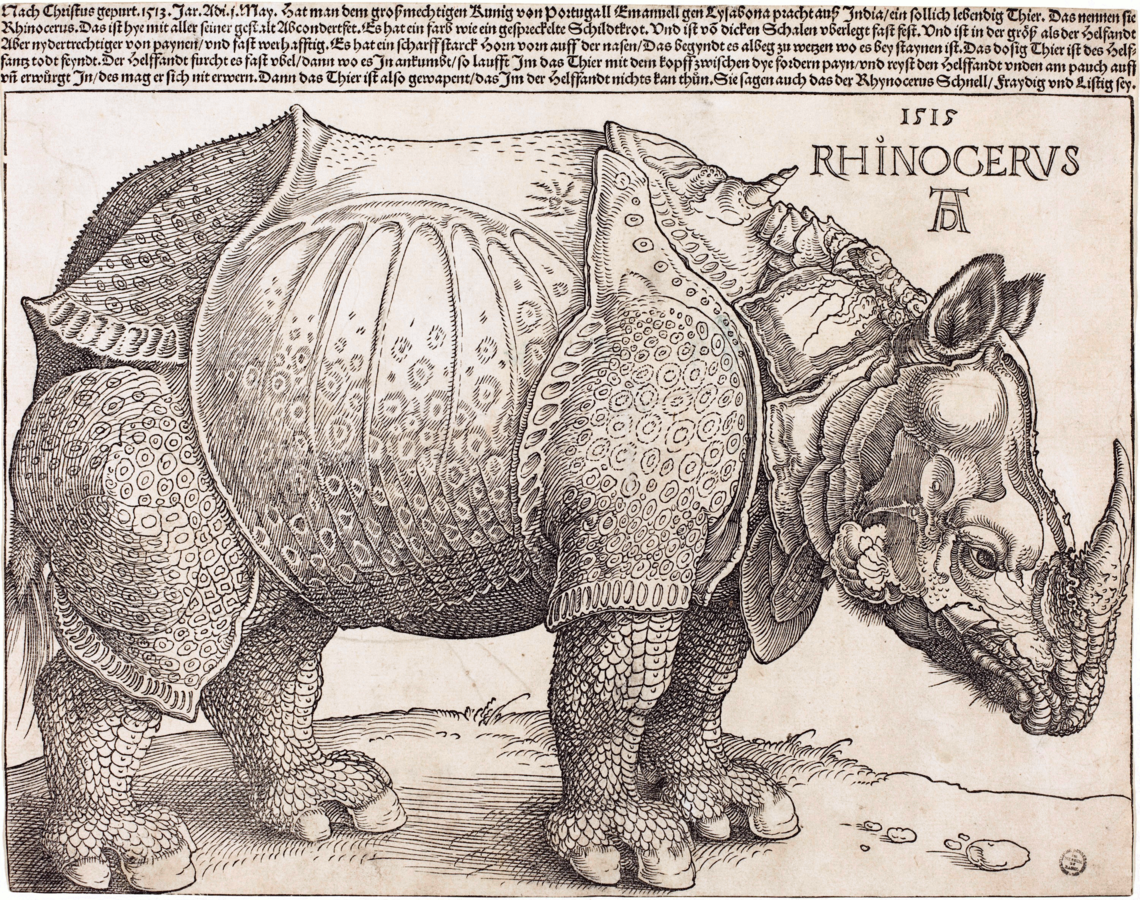}
    \caption{The Rhinoceros, Albrecht Dürer.}
    \label{fig:Durero}
  \end{minipage}
\end{figure}

Modality changes in most cases imply a loss of information even for humans. There are many examples of recreations only from textual descriptions that had little in common with the original images. For example, in medieval times, paintings of animals from textual descriptions were made by painters who had never seen such animals or images of them. The results were in many cases completely different from the real animal \cite{MedievalAnimals1}. One example is shown in Figure \ref{fig:Bestiary} representing elephants. This example is taken from the Rochester bestiary, one of the manuscripts that described and represented exotic animals \cite{Bestiary}.

These inaccurate drawings of animals have in some cases become the pattern to represent them. A well-known example is a woodcut made by German artist Albrecht Dürer of a rhinoceros \cite{ogasawara2021durer}. In 1515, a living rhinoceros was brought to Europe but Dürer never saw the animal, his representation was based on a written description and a sketch. As a result, Dürer's woodcut was not accurate as can be seen in Figure \ref{fig:Durero}. However, Dürer's rhinoceros was copied many times and was considered a true representation of a rhinoceros for more than a century having a large influence on the arts. This example shows how information loss due to modality changes can have a large impact, in this case, one that lasted centuries influencing artists such as Salvador Dalí and more recently AI recreations \cite{birkin2024durer}.

In other cases, we do not have the real object to compare with but only a textual description. This is the case for example of the Tower of Babel \cite{Babel}, the Temple of Solomon \cite{Solomon}, or the Labyrinth of Crete \cite{Crete} among many others. The attempts to recreate those objects have produced very different outcomes depending on the artist and the historical period, implying that a textual description of an object is not enough to uniquely define its representation as an image. In one recent study \cite{merino2024word}, text-to-image AI models are used to recreate an ancient Roman village described by Pliny the Younger in a letter with results that are completely different from previous recreations by architects and artists over time. 

These examples show that information loss is expected as we change from text to image and vice-versa. However, for humans, it seems reasonable to assume that if we start with a description of an exotic animal, another person draws the animal from that description then a second person writes a description from the drawing, a third person draws the animal from the second description, and so on, after several changes we would still have an animal that may be completely different from the original one but still an animal. This is not necessarily the case of modality changes when using AI models as we will see in the results presented in the next sections.

\section{Methods}
\label{Methods}

This section first discusses the recursive process of modality changes to then describe the datasets, models, and experiments as well as the metrics used to estimate the similarity across generations. 

\subsection{Recursive Modality Changes (RMC)}

\begin{figure*}[h]
  \centering
  \includegraphics[scale=0.7]{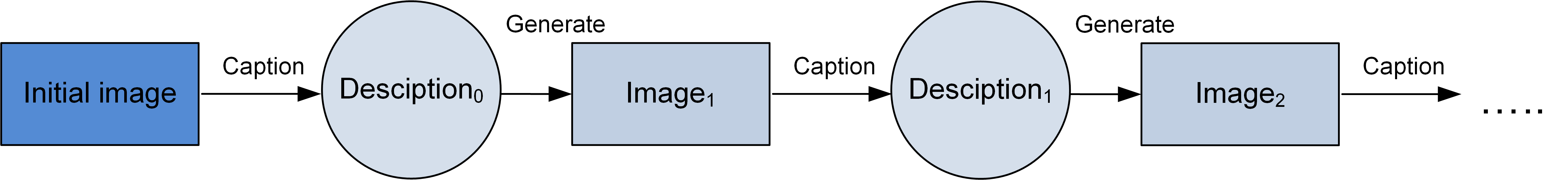}
  \caption{Illustration of the Recursive Multimodal Change (RMC) process.}
  \label{fig:RMC}
\end{figure*}

The process of recursive modality changes is illustrated in Figure \ref{fig:RMC}, it starts with an image that is used as the input to the multimodal AI model to generate a description. Then the description is used to generate a new image using a text-to-image  AI model that is again used for description extraction and so forth. Ideally, the content generated after a few iterations would be similar to the initial one. However, that may not always be the case.

In the context of RMC a trajectory (\textit{t}) is composed of different generations (\textit{g}) that belong to the same original image (image$_0$). Generations $g_i$ or $g_j$ refer to the generation at position \textit{i} or \textit{j}. In this way, generation $g_i$ is composed of image$_{i}$ and description$_{i}$ (extracted from image$_{i}$). This description$_{i}$ is the one used to generate image$_{i+1}$. Generations that are at distance \textit{d} are those that are separated by $|j-i|$ iterations.

\begin{figure}[h]
    \centering
    \includegraphics[scale=0.23]{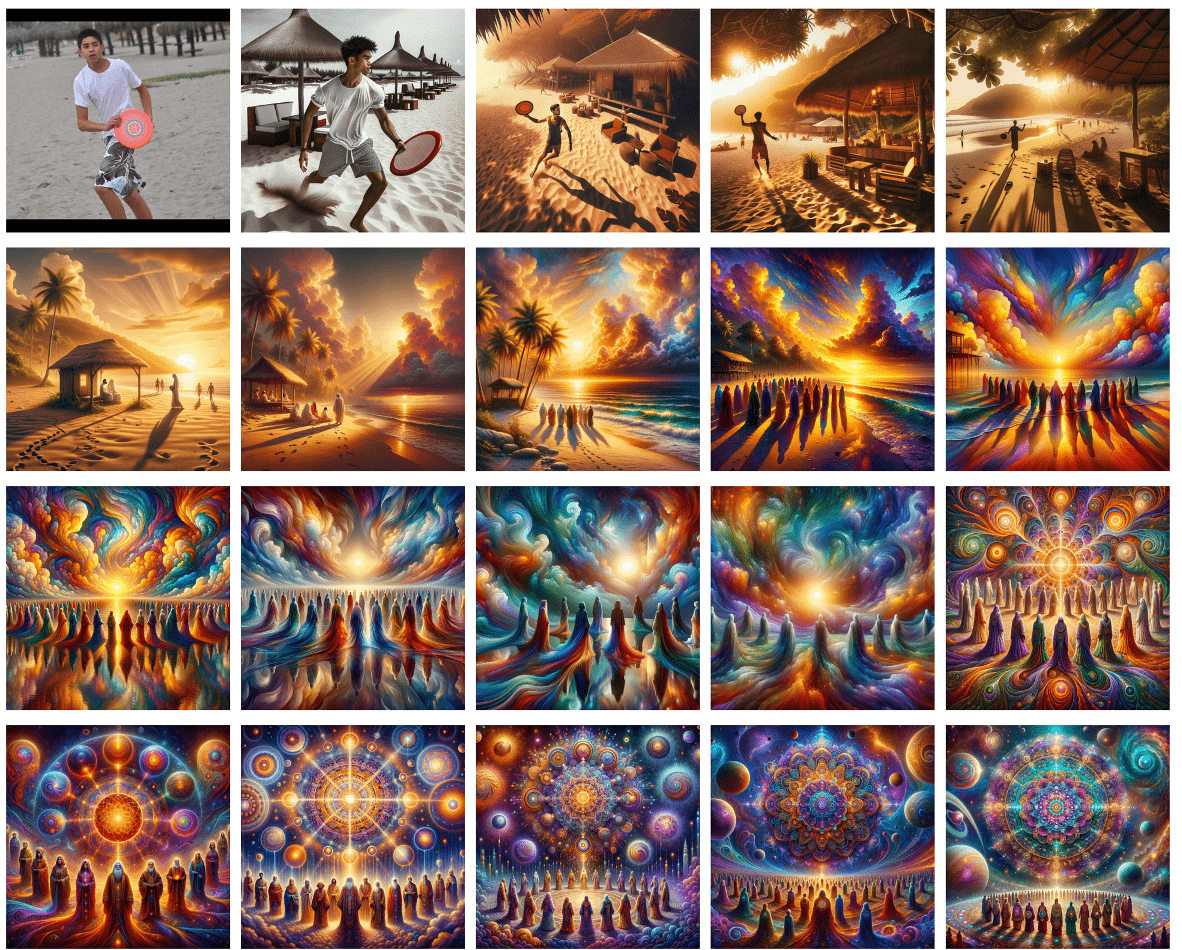}
    \caption{Example of the RMC process (image 103227). The original image is on the top-left and iterations increase to the right and with each row.}
    \label{fig:RMC_example}
\end{figure}

An example of the RMC process is shown in Figure~\ref{fig:RMC_example} when using GPT-4o as the multimodal AI model and DALL-E 3 as the text-to-image AI model.

This trajectory starts with an initial image showing a child playing on the beach with a frisbee. The first description extracted with \textit{GPT-4o} describes the image as follows. \textit{Description$_{0}$: In this image, a young individual is captured mid-movement on a sandy beach, holding a large red frisbee. Dressed in a casual white T-shirt and patterned shorts that showcase a mix of grey, black, and white, they are seemingly engaged in an active beach game. The backdrop features distinctive thatched beach umbrellas and benches, creating a relaxing and tropical atmosphere. The sandy terrain stretches into the distance, with patches of grass adding a hint of greenery to the otherwise neutral-toned landscape. The scene has a candid and energetic vibe, perfectly capturing the essence of beachside recreation.}

The description provided by GPT-4o details the entire image. It not only describes the child but also the setting, clothing style, colors, objects, etc. In the early generations, the model maintains the context of the image. However, the child quickly shifts from being the central focus of the image to just another element, until disappearing (generation 5). From that point on, the loop remains the beach, but starting from generation 7, groups of people begin to appear near the shore, and the sky becomes multicolored with a futuristic style. By generation 10, the beach transforms into a completely different scene where people are wearing long cloaks and forming organized groups around a sky with ethereal interconnected shapes. This leads to the final generation of this example (generation 19), where GPT
4-o describes it as follows. \textit{Description$_{19}$: This captivating image portrays a mystical and cosmic scene that combines spiritual and celestial elements. The centerpiece is an intricate mandala, radiating vibrant colors and exquisite geometric patterns, creating a mesmerizing focal point. Surrounding the mandala are numerous celestial bodies, including planets and swirling galaxies, set against a rich, star-studded background that conveys an infinite, otherworldly landscape. Gathered in a large circular formation beneath the mandala, robed figures stand in reverence, each adorned with ornate, mandala-like symbols on their backs. Their garments are vibrant and elaborately designed, reflecting the swirling hues and patterns of the cosmos above them. Light beams and cosmic energy seem to connect the mandala with the assembly below, suggesting a deep spiritual connection and unity between the celestial and the earthly realms. The overall style is a blend of psychedelic art and visionary imagery, filled with rich, luminous colors, intricate details, and a profound sense of cosmic wonder and spiritual enlightenment.} 

As can be seen in this case, RMC has evolved into a scene completely different from the initial concept, both in the descriptions and in the image.


\subsection{Dataset description and Sample}

The images used in our evaluation were taken from the Common Objects in Context (COCO) dataset~\cite{coco_dataset}, a dataset with more than 330k images containing everyday objects such as people, animals, vehicles, etc., in different contexts. The dataset organizes the images by categories and subcategories, includes 5 brief captions per image, and offers object segmentation on the images. The COCO Dataset is a reference dataset used for training deep neural networks in tasks of image classification, captioning, and object detection.
For this study, 10 images containing people (category ``person'') were randomly selected, ensuring that the images do not contain more than three types of tagged objects to ensure that the input images are primarily focused on people.

Similarly, the study was repeated by choosing the categories ``elephant'', ``toilet'', ``train'', ``apple'' and ``fire-hydrant''. These categories were selected as they allow for the evaluation of the experiment's results in six different contexts: 1) people, 2) other animals that are not people, 3) vehicles, 4) food, and 5) inanimate objects. Additionally, we selected images that meet the requirement that category is one of the central elements by ensuring that they are images with three or fewer tagged objects.

\subsection{Models}

For the development of the study, two different and among the most advanced models in the state of the art to date according to benchmarks have been used: 1) the multimodal GPT-4o that allows extracting prompts from images, and 2) the DALL-E 3 model for generating images through prompts. Additionally, these models were selected because they belong to the same company (OpenAI) and, as of today, GPT-4o does not allow generating images. The API documentation mentions DALL-E 3 as an alternative for generating images. The interaction with the models was done through the OpenAI API, and the model configurations were as follows:

\begin{itemize}
    \item GPT-4o: Model ``gpt-4o-2024-05-13'', with the prompt ``Describe this image respecting the style of the image'' and passing the URL where the image is hosted. The maximum number of tokens to generate is 1,000, and the Fidelity image understanding level is high for capturing all the details of the image. The rest of the parameters were left at default to try to replicate a scenario close to reality. That is, the frequency penalty and presence penalty are 0, with a single generation per prediction, and the temperature and top\_p are 0. No seed was set as the variability within the same image will also be evaluated.

    \item DALL-E 3: Model ``dall-e-3'', with the prompt being the description of each image. Regarding the model parameters, the quality was set to ``hd'' to capture finer details and guarantee consistency across the image, size 1024x1024 (the minimum size of DALL-E 3). For the style of the generated images, both options supported by the API were explored. The ``vivid'' style generates hyper-real and dramatic images, while the ``natural'' style produces less hyper-realistic but more natural images.
\end{itemize}

\subsection{Description of the experiments}

As described above, the generation loop starts with an initial image extracted from the COCO dataset. GPT-4o is used to extract the description of the image, and the description is used to generate a new image with DALL-E 3 (with its default style, i.e., vivid, and high definition [hd] quality). Following this loop, images and descriptions will be analyzed by calculating the differences between different generations, images with descriptions, descriptions with images, and descriptions between generations. For this, we propose the following analysis:

\begin{itemize}
    \item Intra-trajectory analysis. The generation loop is performed obtaining 10 trajectories per category each made of 40 different generations obtained from the images extracted from the COCO dataset. As a result, 2,400 different images are obtained and 49,200 comparisons (820 comparisons per trajectory; all possible combinations of generations without repetition).

    \item Comparison between different categories. This scenario focuses on observing differences based on the category of the input dataset (``person'', ``elephant'', ``toilet'', ``train'', ``apple'' and ``fire-hydrant'').

    \item Comparison of different DALL-E 3 configurations. The same experiments are repeated with DALL-E 3 configured to generate images with the ``natural'' style. This analysis aims to observe if there are differences in the results across the different styles of DALL-E 3.
    

    \item Inter-trajectory analysis. This experiment studies the influence of the number of generations. Instead of comparing the different generations belonging to the same trajectory, a comparison of each generation among the different images is made. The objective is to measure if the model follows any trend regardless of the base image.

\end{itemize}

\subsection{Similarity between generations}

To analyze how RMCs can affect content across generations, we propose the use of image and text similarity metrics. In particular, we use Learned Perceptual Image Patch Similarity (LPIPS) \cite{LPIPS} for images, Bidirectional Encoder Representations from Transformers (BERT) for texts, and Bootstrapping Language-Image Pre-training (BLIP) \cite{blip} to compare images and text. These metrics are based on extracting the embeddings of the items to compare using an AI model and then computing the cosine distance of the embeddings to estimate the similarity. To compare texts, Term Frequency - Inverse Document Frequency (TF-IDF) \cite{TF-IDF} is also used. In this case, a value derived from the frequencies of the words in the text is used as the metric for each word, and the cosine distance on those values is used to estimate similarity.  

\begin{figure*}[h]
  \centering
  \includegraphics[scale=0.6]{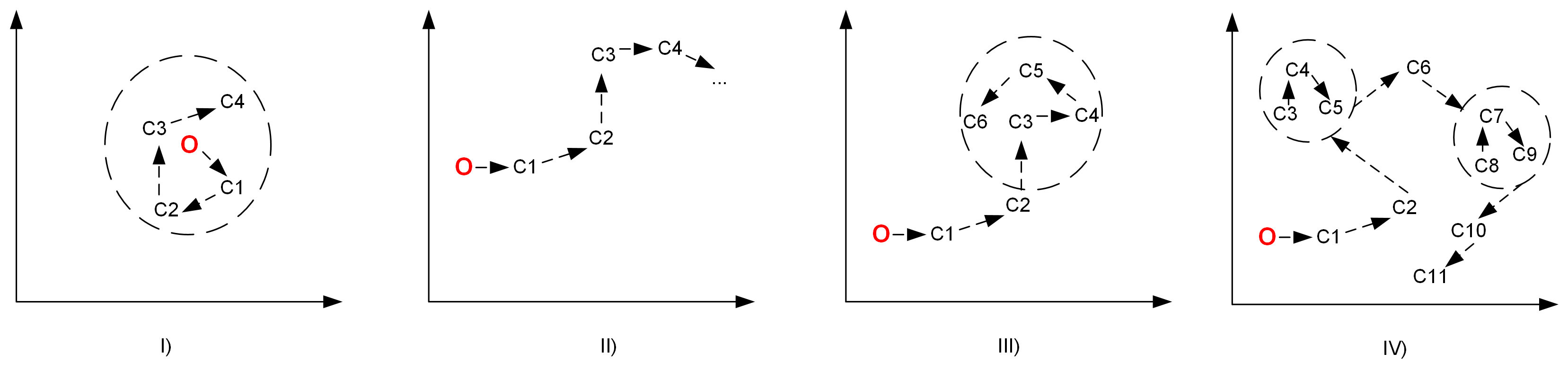}
  \caption{Examples of the potential effects of Recursive Modality Changes on the embeddings of the images/text: I) the embeddings of successive generations (Cx) are close to the embeddings of the initial content (O); II) the embeddings diverge from the original content and traverse the embedding space; III) the embeddings move to another zone in the embedding space and then stabilize there; IV) the embeddings move to another zone, stabilizes for some iterations and then diverges again}
  \label{fig:RMC-embeddings}
\end{figure*}

These metrics enable us to assess if the generations are similar or diverge. This is illustrated in Figure \ref{fig:RMC-embeddings} using two-dimensional embeddings for visualization. The first plot shows a stable RMC process on which embeddings remain close to the original content (I), the second plot a divergent process (II), the third plot the case in which there is a movement to another zone of the embeddings' space on which the changes converge (III), and the last plot the embeddings converge during some iterations before diverging again (IV).

\begin{figure}[h]
  \centering
  \includegraphics[scale=0.10]{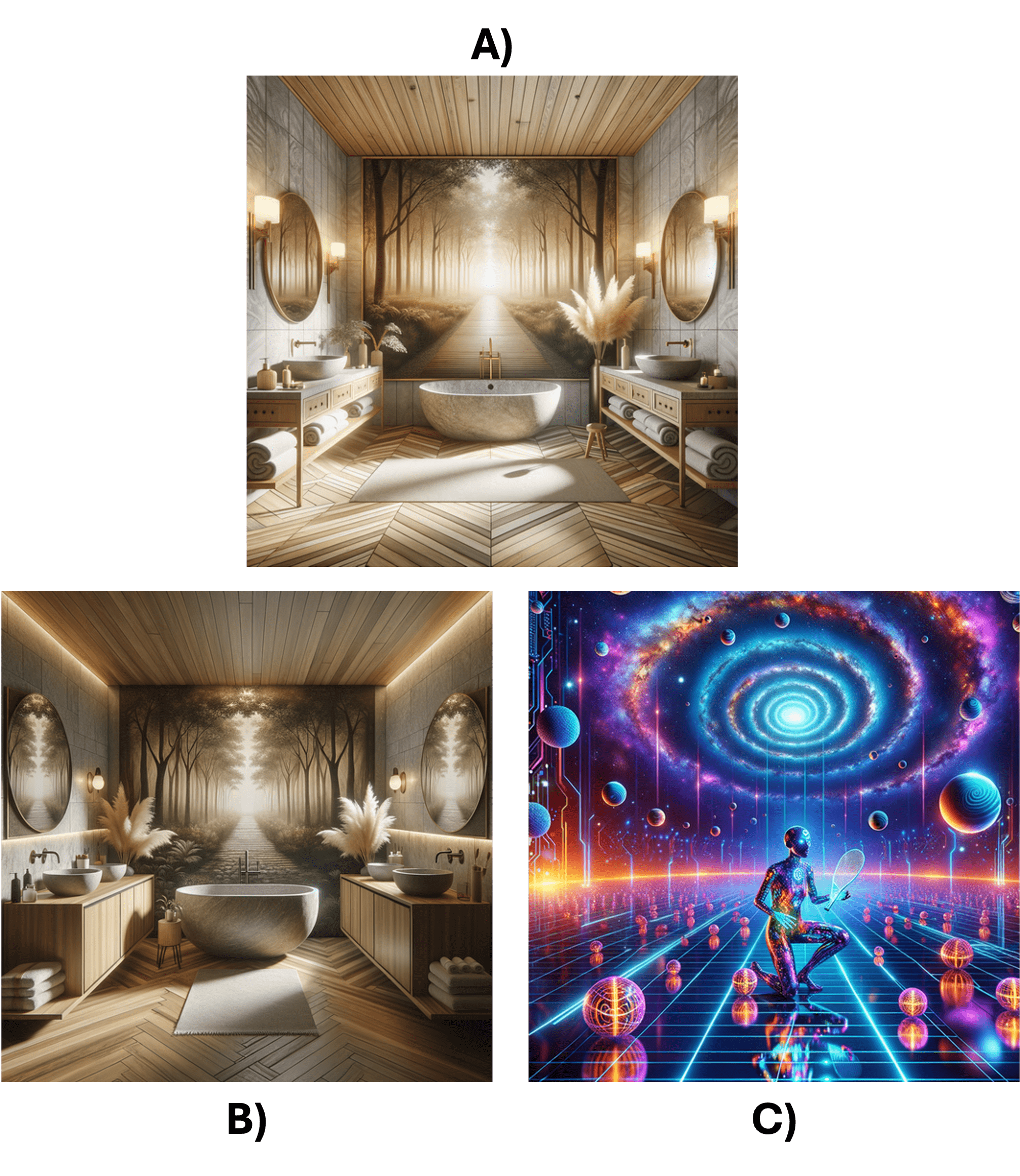}
  \caption{Reference images generated by DALL-E 3 (the complete description can be found on the Zenodo Dataset). A) (image 66859 - Generation 18) \textit{This image showcases a luxurious, nature-inspired bathroom adorned with refined details...}; B) (image 66859 - Generation 17) \textit{This image depicts a luxurious, nature-inspired bathroom interior with a serene and sophisticated ambiance. The focal point of the space is a stunning mural of a sunlit forest path, creating an immersive and tranquil atmosphere...}; C) (image 251144 - Generation 39)\textit{This image is a vibrant, futuristic depiction that seamlessly blends elements of science fiction and surrealism...}
  }
  \label{fig:references}
\end{figure}

As an example and reference for the rest of the article, we obtained these metrics by comparing a reference image (A) with another very similar image (B) and with a completely different image (C). All images were generated with Dall-E 3, and their descriptions were extracted using GPT-4o. Figure~\ref{fig:references} contains the three images, and Table~\ref{tab:metrics} summarizes the comparison metrics. It can be observed that the similar images (A-B) show a high level of similarity between the images (LPIPS), their descriptions (BERT and TF-IDF), as well as between the image and description (BLIP-1) and vice versa (BLIP-2). In contrast, the completely different images (A-C), all metrics reflect lower values in all the cases. The comparison of an image with itself is maximum (i.e., 1) in each case, except for BLIP, since it measures the similarity between the description and the image. It should be noted that in the case of LPIPS for images generated with DALL-E, the value is quite small for two very similar images, meaning that small changes in the images produce large changes in LPIPS. In the case of very different images, this value is indeed very small. We will use these metrics in relative terms; that is, images A-B are more similar to each other than images A-C.

\begin{table}[]
\caption{Similarity metrics when comparing image A) [Image 66859 - Generation 18] with B) [Image 66859 - Generation 17] and C) [Image 251144 - Generation 39] }
\centering
\begin{tabular}{cc|c|c|c|}
\cline{3-5}
\multicolumn{1}{l}{}                     &        & A & B & C \\ \hline
\multicolumn{1}{|c|}{\multirow{5}{*}{A}} & LPIPS  & 1.00  & 0.51 & 0.20   \\ \cline{2-5} 
\multicolumn{1}{|c|}{}                   & TF-IDF & 1.00  & 0.72 & 0.50   \\ \cline{2-5} 
\multicolumn{1}{|c|}{}                   & BERT   & 1.00  & 0.97  & 0.86  \\ \cline{2-5} 
\multicolumn{1}{|c|}{}                   & BLIP-1{\scriptsize$^{1}$} & 0.50  & 0.49  & 0.23  \\ \cline{2-5} 
\multicolumn{1}{|c|}{}                   & BLIP-2{\scriptsize$^{2}$} & 0.50  & 0.50  & 0.18  \\ \hline
\multicolumn{5}{l}{\begin{tabular}[c]{@{}l@{}}
{\scriptsize$^{1}$}BLIP image A - description A/B/C\\ 
{\scriptsize$^{2}$}BLIP description A - image A/B/C 
\end{tabular}
}

\end{tabular}
\label{tab:metrics}
\end{table}

\section{Results}
\label{Results}

This section presents and discusses the results of the different experiments, all the images and texts produced in the experiments are available in a Zenodo public repository\footnote{Dataset: \url{https://zenodo.org/doi/10.5281/zenodo.13362646}}.

\subsection{Intra-trajectory analysis}

In this first analysis, the trajectories will be analyzed to observe the evolution across generations through the applied multimodal change cycle.

\begin{figure}[h]
  \centering
  \includegraphics[scale=0.6]{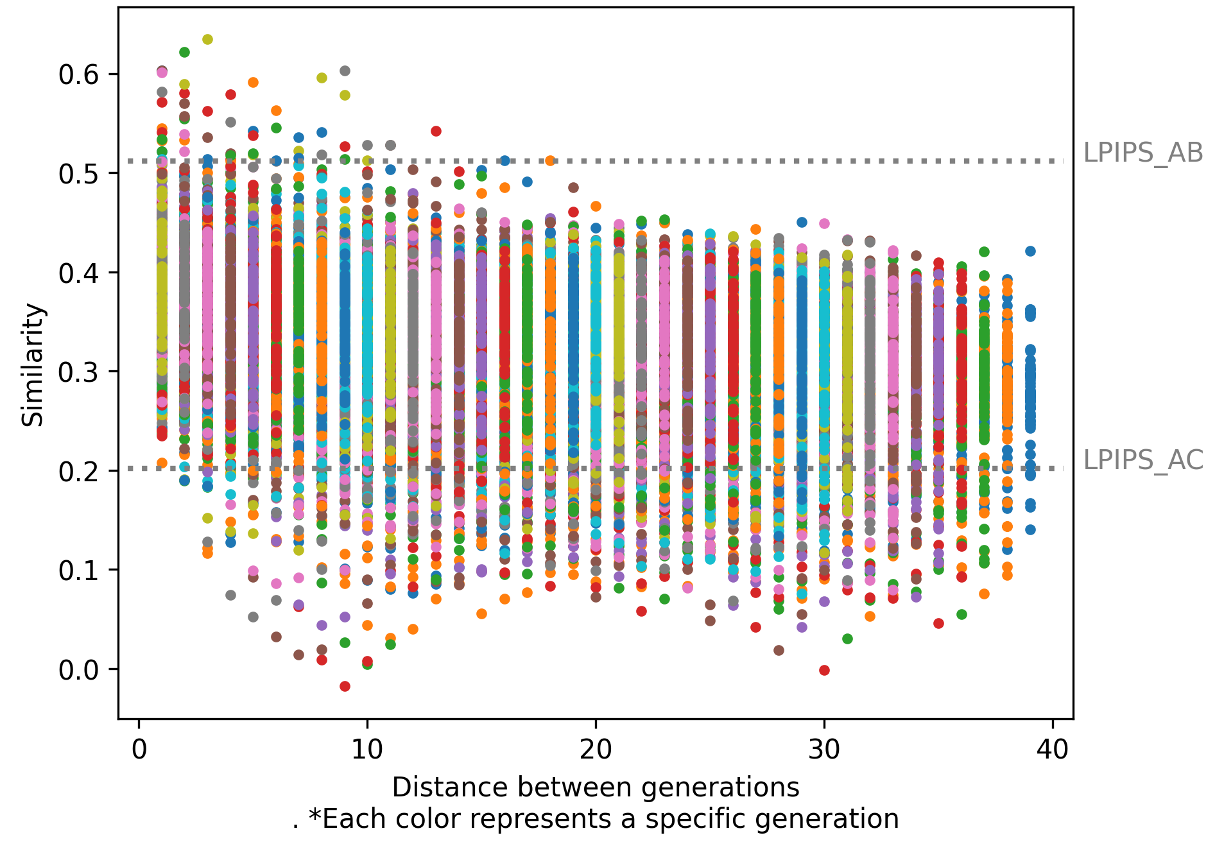}
  \caption{LPIPS similarity of each generation in comparison with the other generations. Similarity is computed as $d=i-j$ being $i$ and $j$ the number of generations. $ \forall i,j \in [0, 39]; i > j$.}
  \label{fig:VGG_all}
\end{figure}

In this analysis, a comparison is made between images that belong to the same trajectory. The data shows that as the distance between generations increases, the difference between images also increases, with the minimum LPIPS similarity for d=39 (mean = 0.27, median = 0.28, std = 0.05) while at shorter distances it increases, with the maximum LPIPS for d=1 (mean = 0.38, median = 0.38, std = 0.05). Figure~\ref{fig:VGG_all} represents the LPIPS similarity for all the combinations. The same results are observed when comparing the descriptions, where a minimum cosine similarity is obtained for d=39 for both TF-IDF (mean = 0.53, median = 0.53, std = 0.08) and BERT (mean = 0.88, median = 0.89, std = 0.04); and a maximum for d=1 for both TF-IDF (mean = 0.68, median = 0.69, std = 0.07) and BERT (mean = 0.95, median = 0.96, std = 0.02).

These results show how the differences both in descriptions and in images increase as more distant generations are compared.

\begin{figure}[h]
  \centering
  \begin{minipage}{0.45\textwidth}
    \centering
    \includegraphics[scale=0.5]{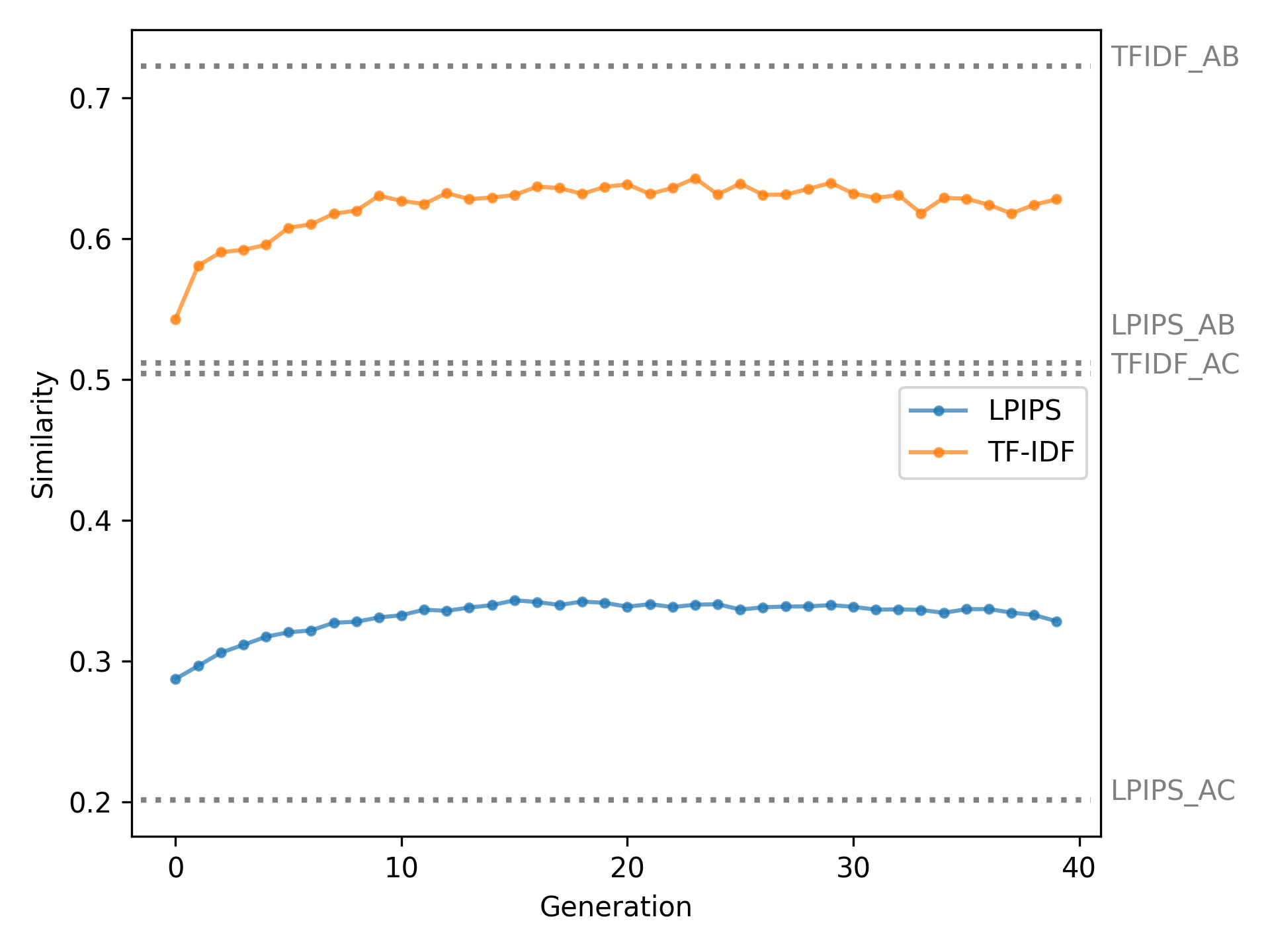}
    \caption{LPIPS mean similarity and TF-IDF mean similarity of each generation in comparison with the other generations with the one of all the dataset.}
    \label{fig:lpips_tf-idf_mean_generation}
  \end{minipage}\hfill
  \begin{minipage}{0.45\textwidth}
    \centering
    \includegraphics[scale=0.38]{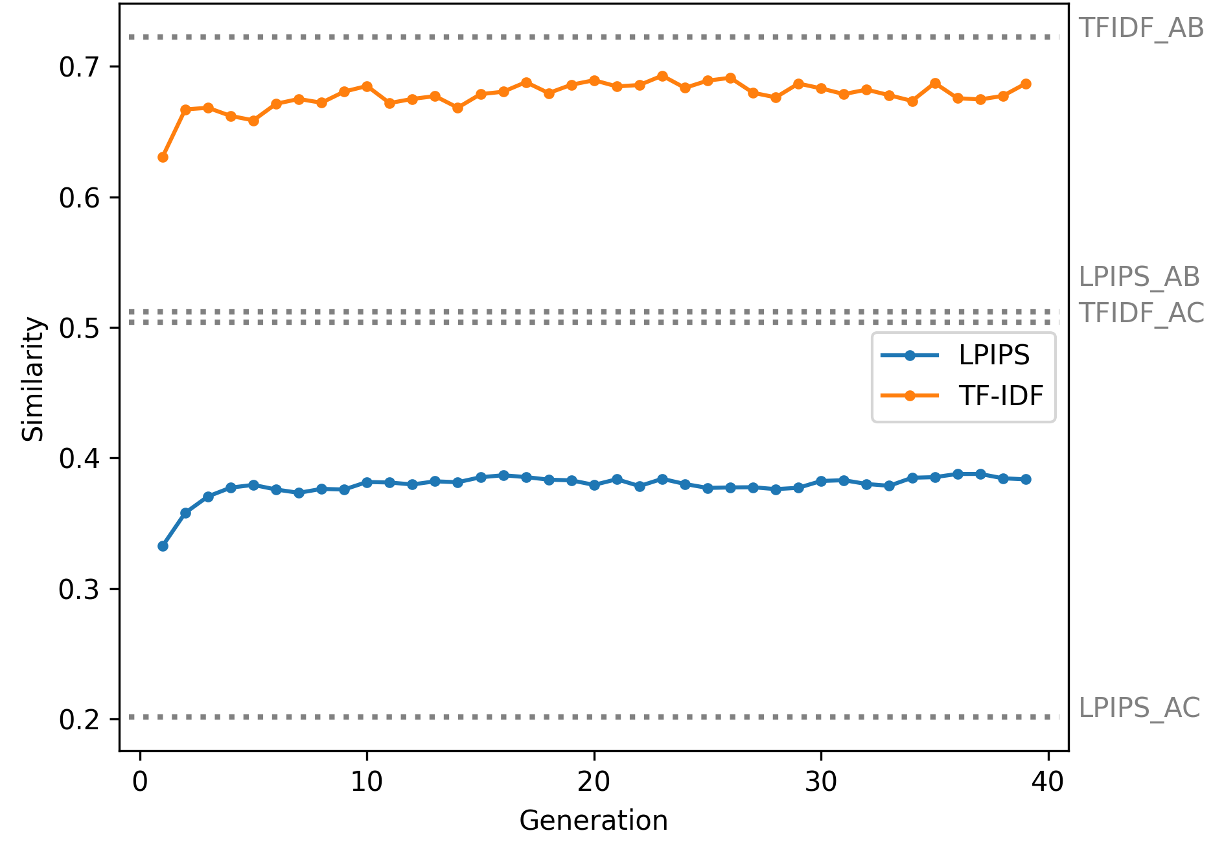}
    \caption{LPIPS mean similarity and TF-IDF mean similarity of each generation in comparison with the previous generation.}
    \label{fig:trajectories_lpips_tf-idf}
  \end{minipage}
\end{figure}

In the same way, the effect of the models can be studied based on the generation number instead of the distance between generations. This allows for the comparison of generations and the study of which generations are most affected by the model. The average similarity between images measured by LPIPS / TF-IDF across all generations is 0.33 / 0.62 (median = 0.34 / 0.63, std = 0.06 / 0.08). Figure~\ref{fig:lpips_tf-idf_mean_generation} shows the average results of these measurements based on the generation compared to the other generations of their trajectory. The results show how the first generations are the most different compared to the rest. That is, degeneration tends to diverge from the initial image, thus discarding hypothesis I). However, from generation 10 onwards, it remains more or less constant with certain peaks both ascending and descending.

To analyze whether the system is in II), III), or IV), we can compare how the images evolve concerning the previous generation. In case III), it should be observed that the similarity would begin to increase as the generations progress. Figure~\ref{fig:trajectories_lpips_tf-idf} shows that this does not happen; it rather seems that there is a lower similarity in the first generations, but in the rest of the generations, it remains stable. This demonstrates how the model does not tend to converge in the trajectories (at least after 40 generations).

This measure, however, does not allow discerning whether the system is in state II) or IV) since it may happen that in each trajectory, the cycles have different durations, which would affect the average results. Therefore, it is proposed that the convergence zone be defined as the one in which N consecutive comparisons of generations (a cycle of duration N contains N+1 consecutive images) remain over the median similarity of LPIPS. The median is used instead of the mean to avoid outliers.

By setting $N = 4$, the results show that the average number of cycles per trajectory is 1.73 in each trajectory, with an average cycle duration of 6.21 (median = 5, std = 2.87). At least one cycle is detected in 95.0\% of the trajectories, with 36.7\% of the trajectories having one cycle, 38.3\% having two cycles, and 20.0\% having three cycles.

\begin{figure}[h]
  \centering
  \includegraphics[scale=0.1]{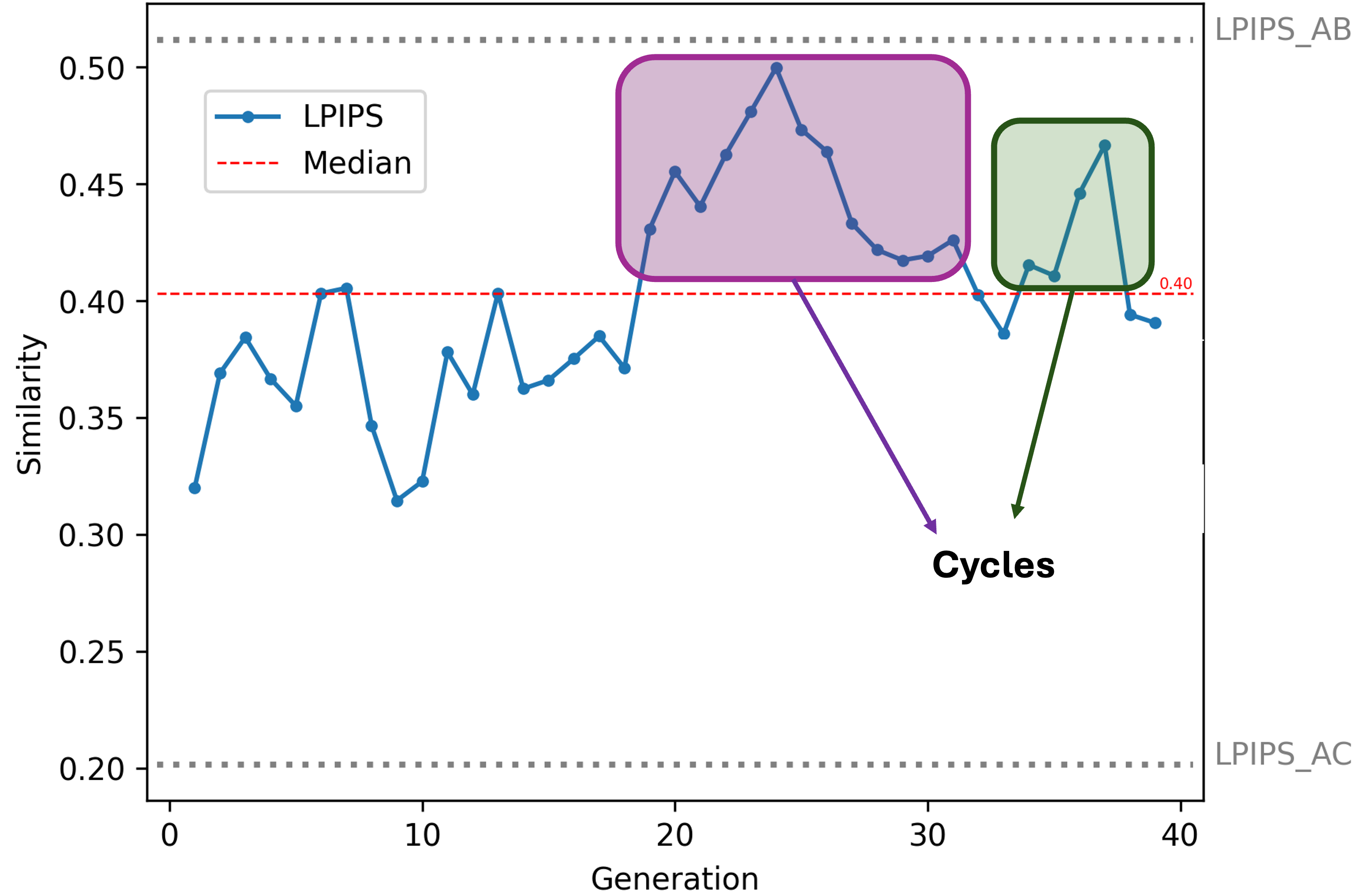}
  \caption{LPIPS similarity with the previous generation of a trajectory with two detected cycles (image 87144).}
  \label{fig:trajectories_lpips_87144_vivid}
\end{figure}

The model does not tend to converge (at least within 40 generations), because we have only detected a few trajectories with cycles in the last generations. For example, Figure~\ref{fig:trajectories_lpips_87144_vivid} shows a sequence in which 2 cycles have been detected, one from generation 18 to 31 (in purple) and another from 33 to 37 (in green).

\begin{figure}[h]
  \centering
  \includegraphics[scale=0.18]{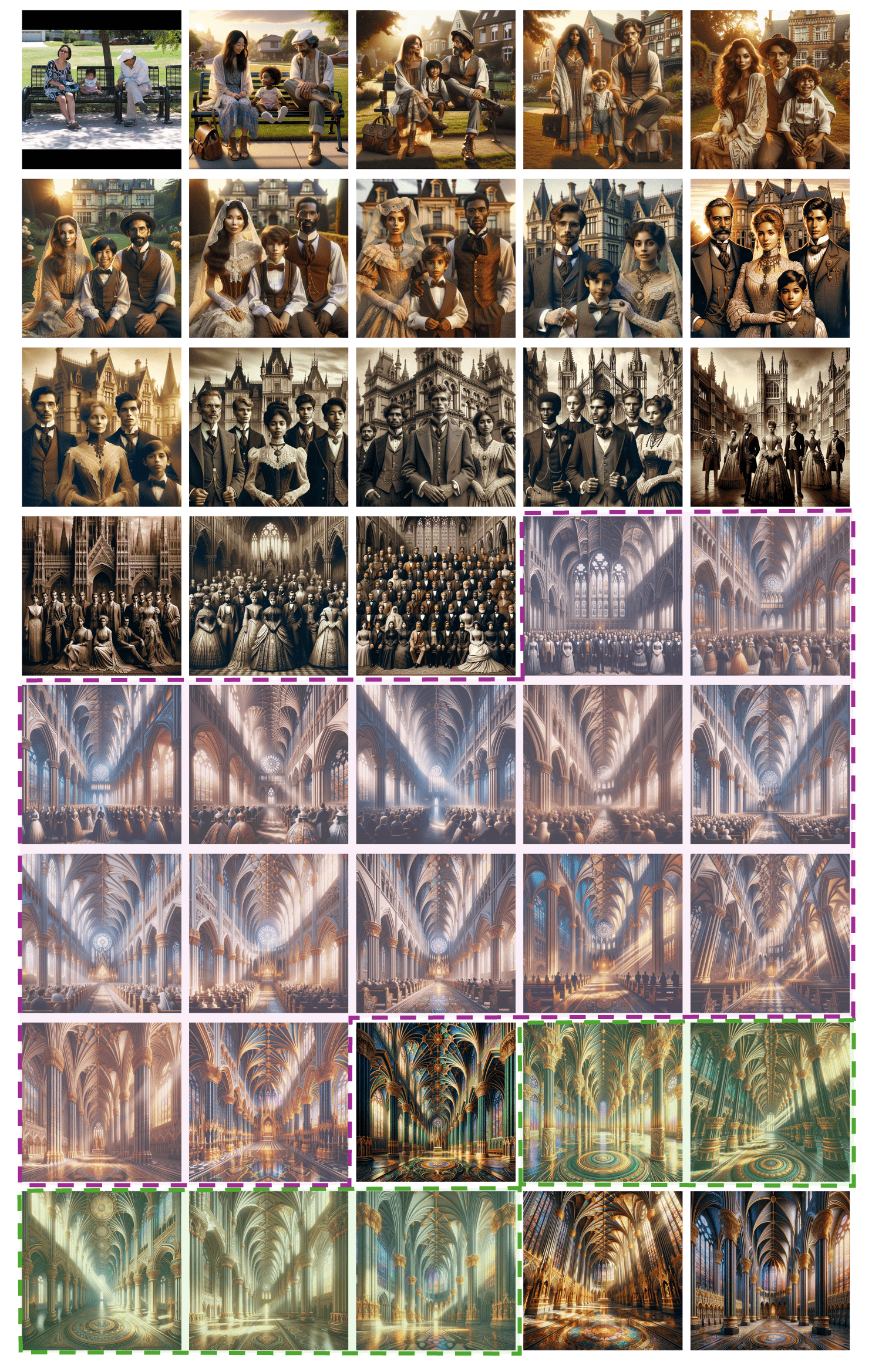}
  \caption{Grid of a trajectory with two detected cycles (iterations in the loops marked with borders in purple and green). The original image is on the top-left and iterations increase to the right with each row (image 87144).}
  \label{fig:grid_87144_vivid}
\end{figure}

This is also observed in the grid of the trajectory (Figure~\ref{fig:grid_87144_vivid}), where it can be seen that in the detected cycles, the images correspond to churches that undergo few modifications except in generation 32, where it changes to a more futuristic style.

The previous results show how the multimodal change process tends to diverge from the initial image. Another alternative to measuring how much it diverges is to detect whether the category of the image is maintained. For example, in a dataset with images of people, check in how many generations of people appear. For this, GPT-4o has been used as a classifier by asking the model if the label (e.g., person) appears in the image. Previous studies have shown that GPT-4o possesses image processing capabilities that enable tasks such as object detection \cite{xin2024dart}. As a sanity check, GPT-4o has been evaluated with the initial image for which we know the ground truth (labels provided by the COCO dataset), resulting in a 100\% accuracy rate in the 60 images tested.

The results show that the initial category is maintained in 64.2\% of the images. However, this percentage is much higher in the early generations (100\% in generation 0; 68.3\% in generation 10; 50.0\% in generation 20; 48.3\% in generation 30; and 40.0\% in generation 39), with the most significant drop occurring in the first 10 generations. These results align with those observed in the previous generation analysis, where the lowest similarity was measured in the early generations compared to the others, and in comparison with the previous generation.

Finally, it is of interest to analyze how text-to-image and image-to-text contribute to the degradation as the GPT-4o model extracts the description from an input image and DALL-E generates an image from a given description. As BLIP allows measuring the similarity between a description and its corresponding image, we have compared image$_{i}$ with description$_{i}$ extracted from that image$_{i}$, and image$_{j+1}$ with description$_{j}$ (the one used to generate the image$_{j+1}$). The goal is to observe which of the two modality changes has a larger impact. The results show that the extraction of the description has the highest similarity with a mean BLIP score of 0.46 (median = 0.47, std = 0.05) compared to the image generation phase (mean = 0.40, median = 0.41, std = 0.08). In other words, it seems that image generation contributes slightly more to instability.

\subsection{Comparison between different categories}

The objective of this experiment is to measure whether the type of image influences the multimodal loop. To achieve this, we have compared the categories of selected images (people, elephants, toilets, trains, apples, and fire hydrants).

\begin{figure}[h]
  \centering
  \begin{minipage}{0.45\textwidth}
    \centering
    \includegraphics[scale=0.28]{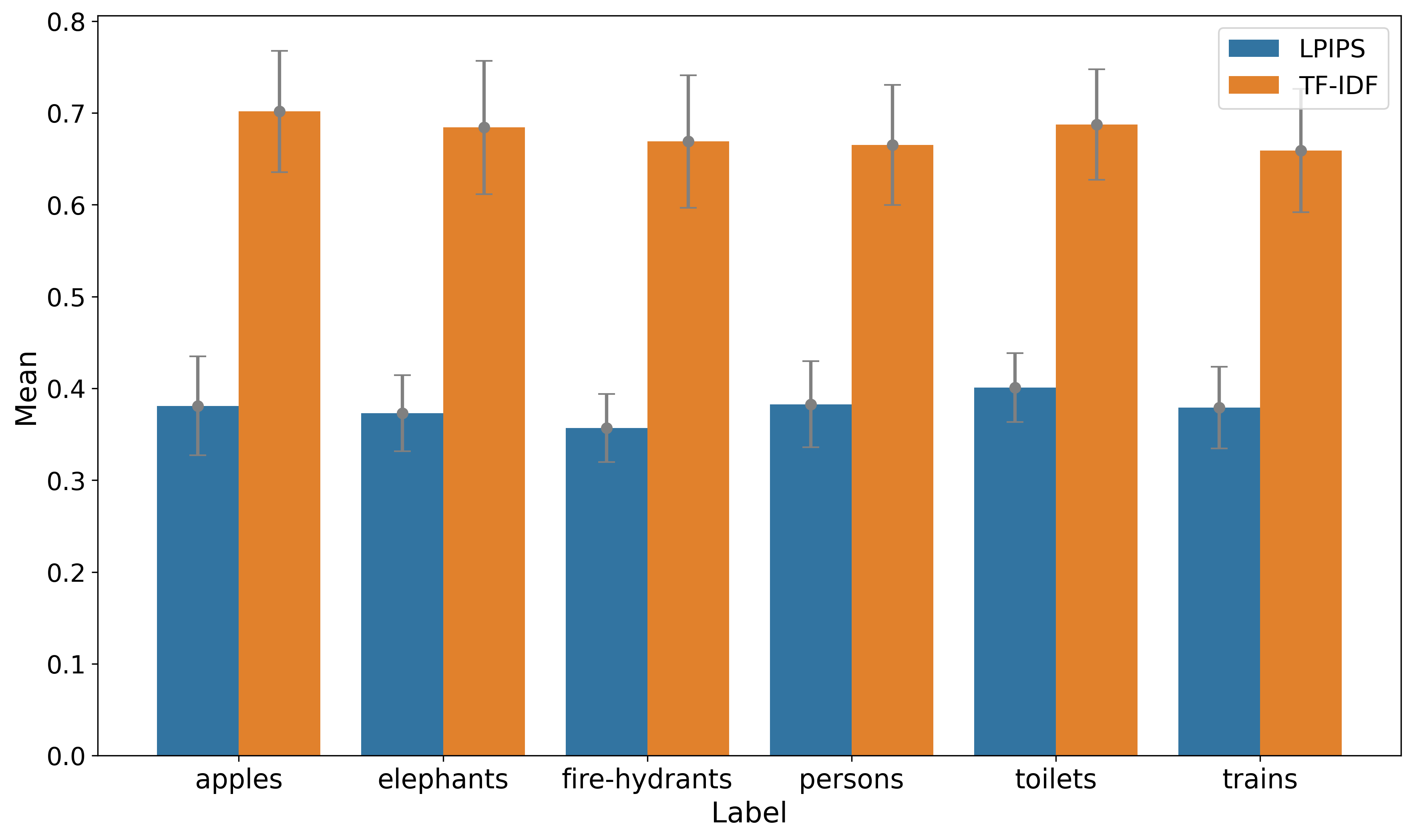}
    \caption{LPIPS mean similarity and TF-IDF mean similarity of each category. Comparison of consecutive generations (d=1).}
    \label{fig:barplot_trajectories}
  \end{minipage}\hfill
  \begin{minipage}{0.45\textwidth}
    \centering
    \includegraphics[scale=0.18]{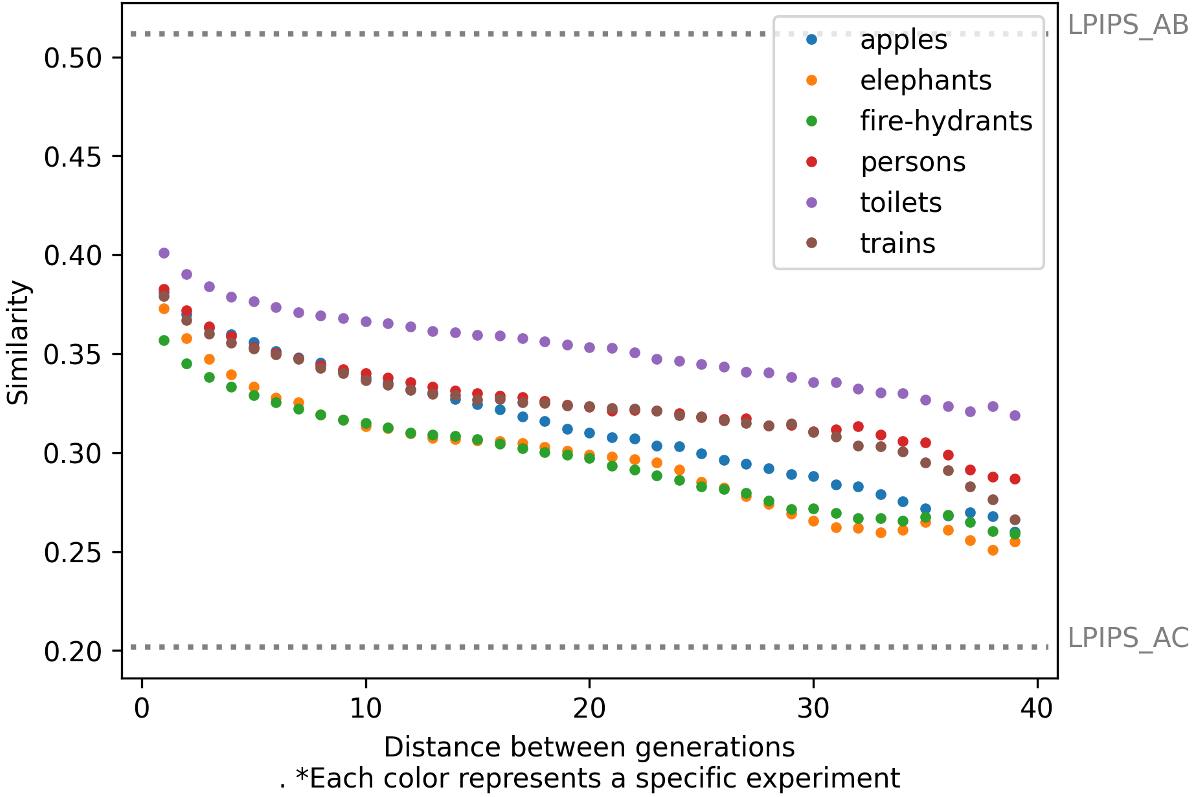}
    \caption{LPIPS similarity (mean) of each category. Comparison of all generations.}
    \label{fig:labels_VGG_mean}
  \end{minipage}
\end{figure}

Figure\ref{fig:barplot_trajectories} shows few differences regarding similarity metrics, with all categories having similarity around 0.4 (LPIPS) and 0.7 (TF-IDF). The variable ``toilets'' is the most stable, while ``fire-hydrants'' and ``elephants'' are the least stable. This can also be seen in the comparison based on distances of generations (Figure~\ref{fig:labels_VGG_mean}), where the category ``toilets'' is the most stable across all the distances and ``elephants'' and ``fire-hydrants'' the least.

Regarding the presence of cycles, Table~\ref{tab:cycles} summarizes the number of cycles and their duration for each category. No significant differences in the number of cycles are observed, with all categories having an average of between 1.6 and 2.0 cycles, with the most common being either 1 or 2 cycles. As for the duration, the most common range is between 4 and 5 generations, although longer cycles (lasting 10 or more generations) have been found in all categories.

\begin{table}
    \caption{Number and duration of cycles in each trajectory per category }
    \centering
    \begin{tabular}{|c|p{3cm}|p{3cm}|} 
    \hline
 category& cycles \newline (mean; mode; std; max)& duration cycles \newline (mean; mode; std; max)\\ \hline
persons &  1.8; 1|2; 0.79; 3 &  6.4; 5; 3.8; 19 \\ \hline
apples &  1.9; 1; 0.88; 3 &  6.6; 4; 3.7; 18  \\ \hline
elephants &  2.0; 2; 0.94; 3 &  5.9; 4; 2.3; 12  \\ \hline
fire-hydrants &  1.5; 1|2; 0.85; 3 &  6.5; 4; 2.5; 10  \\ \hline
toilets &  1.6; 1; 0.97; 3 &  5.8; 5; 2.0; 10  \\ \hline
trains&  1.6; 1; 0.70; 3 &  6.1; 4; 2.5; 12 \\ \hline
total&  1.73; 1; 0.70; 3 &  6.2; 4; 2.9; 19 \\ \hline
    \end{tabular}

    \label{tab:cycles}
\end{table}

\begin{figure}[h]
  \centering

  \begin{minipage}{0.45\textwidth}
    \centering
    \includegraphics[scale=0.33]{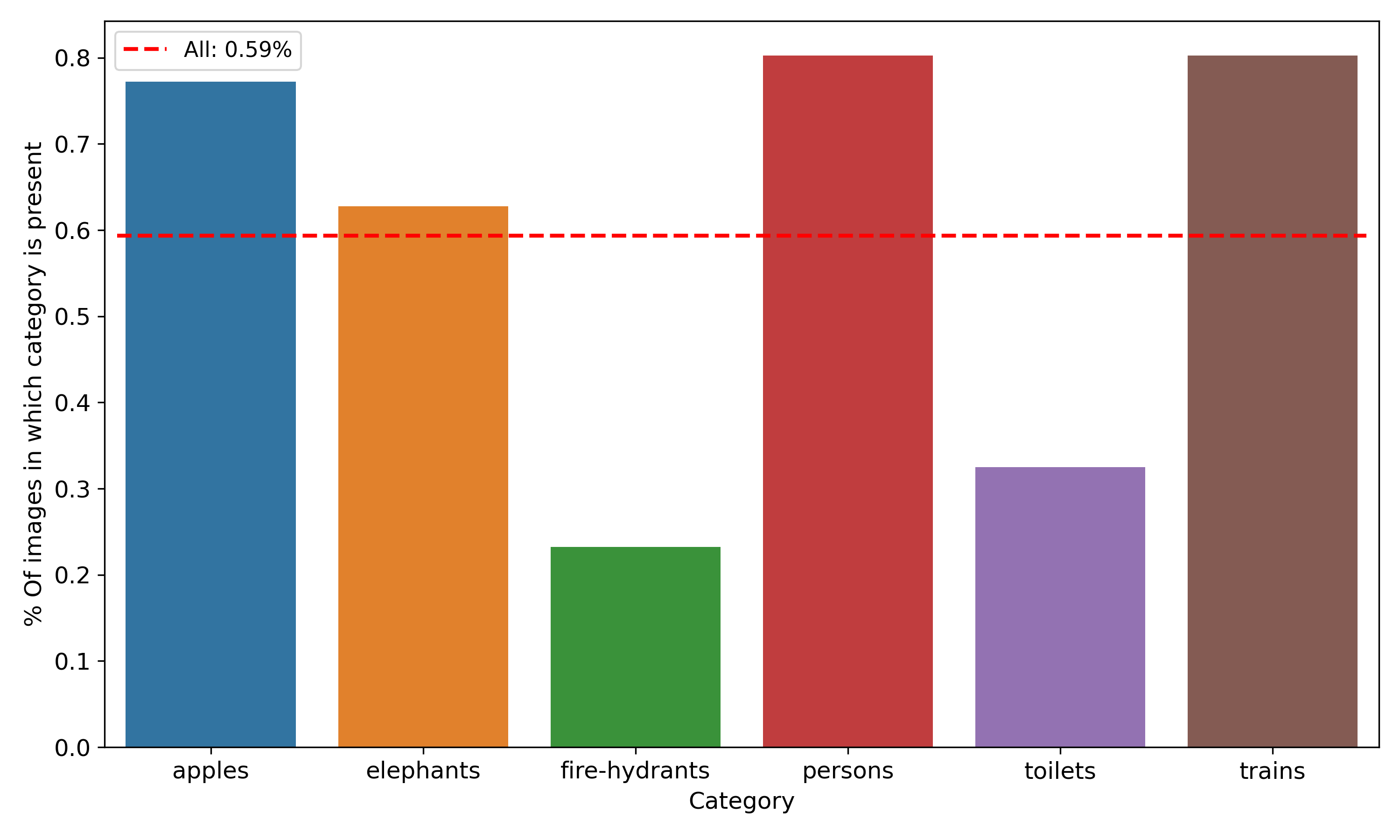}
    \caption{Presence of categories in the images in all the generations.}
    \label{fig:labels_in_image_grouped}
  \end{minipage}\hfill
\begin{minipage}{0.45\textwidth}
    \centering
    \includegraphics[scale=0.09]{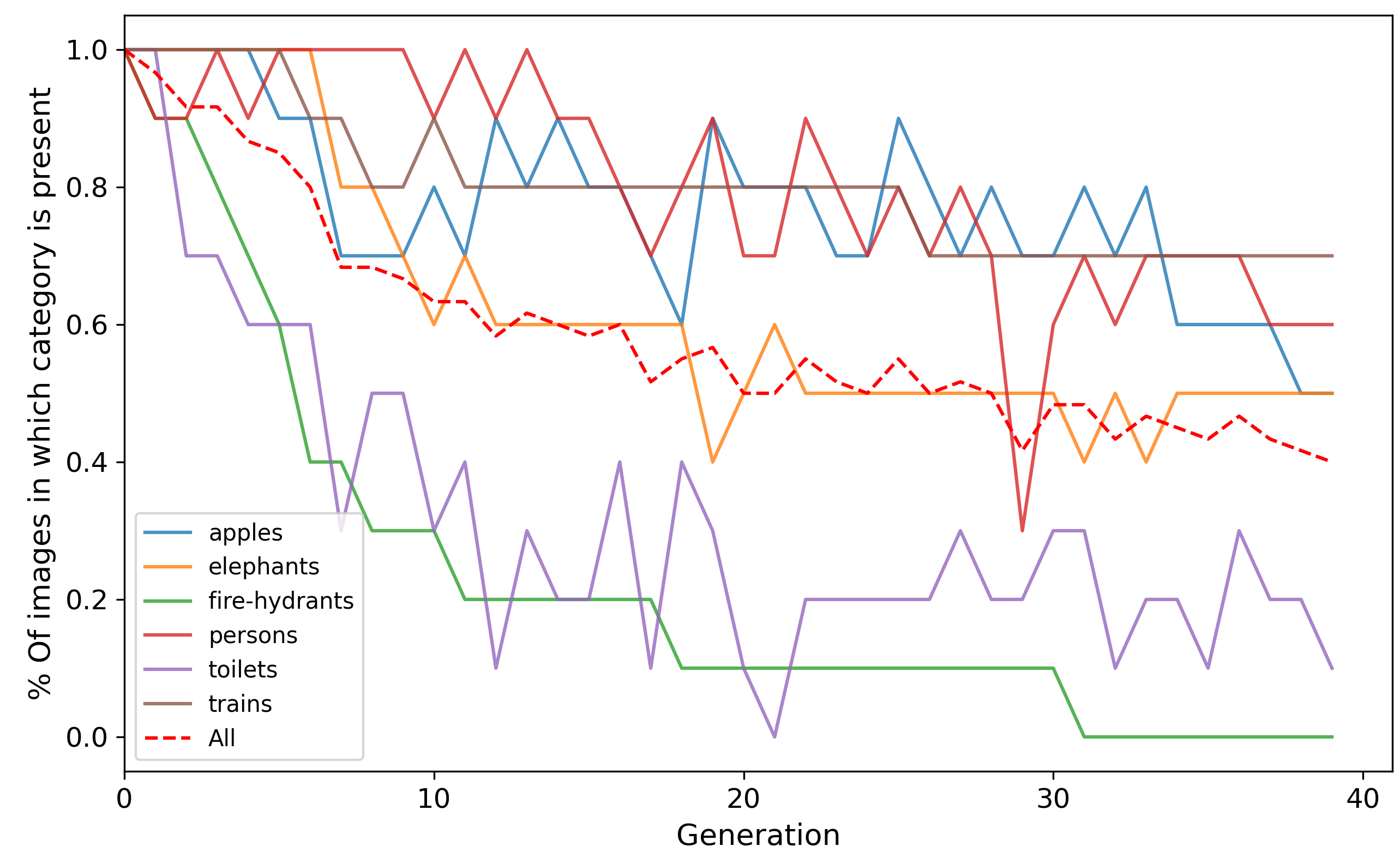}
    \caption{Evolution of presence of categories \\ in the images.}
    \label{fig:labels_in_image_grouped_generations}
  \end{minipage}
  
\end{figure}

Differences in the persistence of categories in the images are indeed observed. Figure~\ref{fig:labels_in_image_grouped} shows how the category persists in around 80\% of the generations for people, trains, and apples, 66.6\% for elephants, 51\% for toilets, and 28\% for fire hydrants. The case of the toilet is particularly interesting, as it showed the highest stability in terms of similarity (LPIPS and TF-IDF). One possible reason is that in the case of toilets, the model tends to maintain the scene (bathroom) without including the toilet, causing metrics that simulate human perception, like LPIPS, to not detect significant changes in the images. Figure~\ref{fig:labels_in_image_grouped_generations} also shows that from generation 10 onward, the presence of toilets drops below 40\%, a trend also observed with fire hydrants but not with the other categories.

\subsection{Comparison of different DALL-E 3 configurations}

One factor to consider when analyzing the multimodality loop is the configuration of the image generation model. We tested an alternative configuration to the default one, using the hyper-realistic style (``natural'').

The analysis was conducted in two distinct phases. In the first phase, 10 different trajectories were generated using 10 images from the ``person'' category. The results showed that for the natural style, the similarity values for LPIPS and TF-IDF were lower (median of 0.27 for LPIPS, and median of 0.59 for TF-IDF) compared to the vivid style (0.34 for LPIPS and 0.60 for TF-IDF).

In the second phase, the aim was to reduce the sample variability by generating 10 trajectories for each style from a single image in the ``person'' category. In this case, the natural style showed a median of 0.19 for LPIPS and 0.56 for TF-IDF, while the vivid style had a median of 0.31 and 0.59, respectively.

Both results prove that vivid style exhibits higher similarity compared to the natural style across all evaluated metrics. This difference is particularly notable in the LPIPS similarity from the second experiment, where the same initial image was used, with median values of 0.31 for vivid versus 0.19 for natural.

\begin{figure}[h]
  \centering
  \includegraphics[scale=0.08]{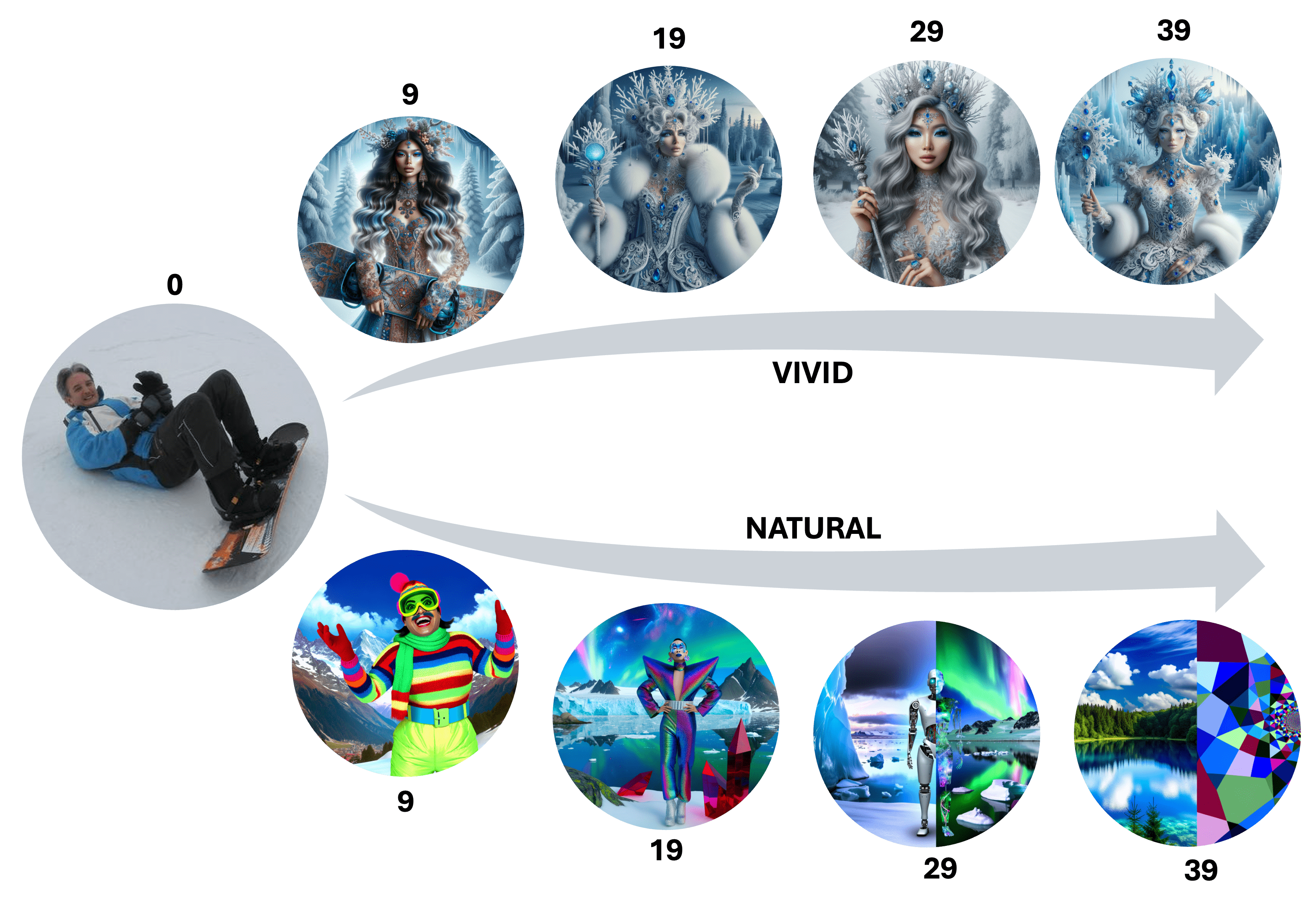}
  \caption{Example of trajectories with vivid and natural styles.}
  \label{fig:styles}
\end{figure}

The greater similarity observed with the vivid style suggests higher stability in the generated images, as evidenced by lower variations in LPIPS and TF-IDF similarity metrics. This may indicate that the vivid style produces more consistent and predictable results compared to the natural style. For example, this can be observed in Figure~\ref{fig:styles} which represents two trajectories of 40 generations generated with both styles.

Although this study focuses on evaluating the default configuration of the models. These results open the door to analyzing different model configurations, including the natural mode, and examining how they affect the multimodality loop.

\subsection{Inter-trajectory analysis}
\begin{figure}[h]
  \centering
  \includegraphics[scale=0.3]{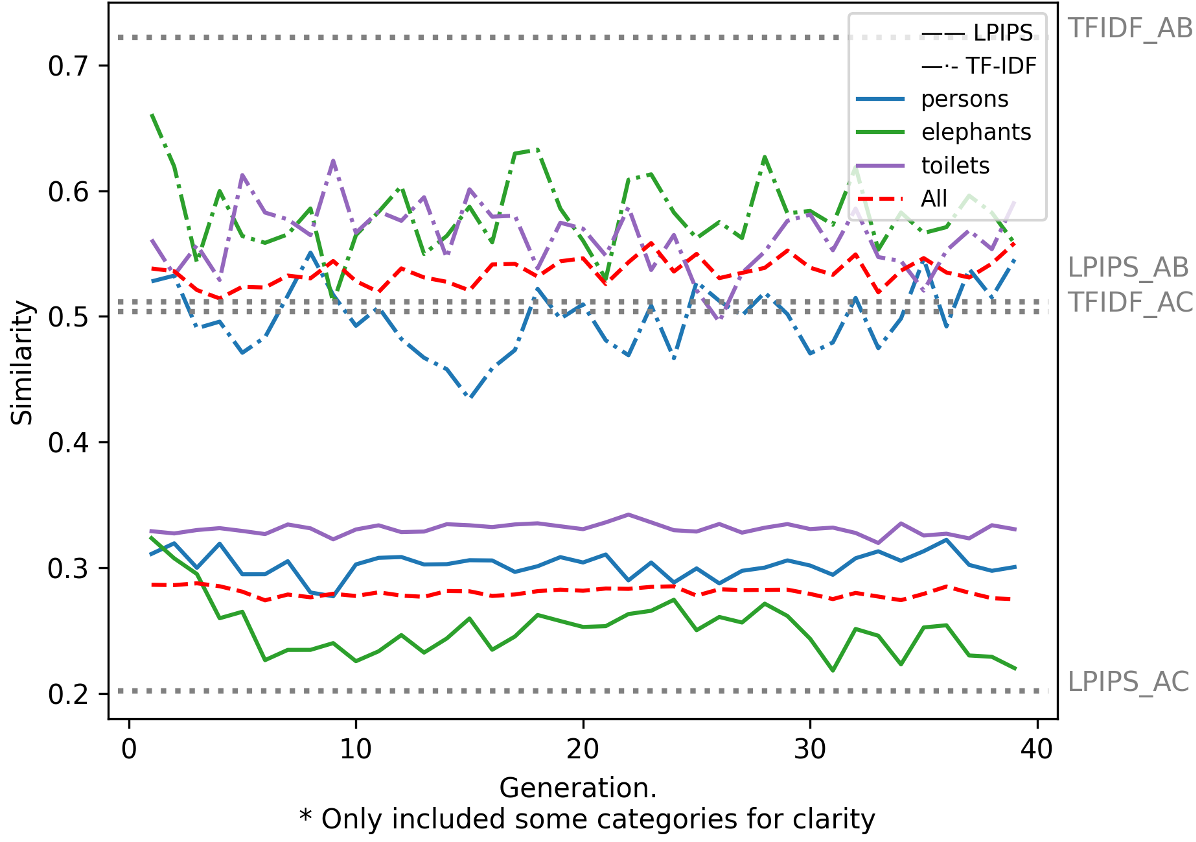}
  \caption{Inter-trajectories LPIPS and TF-IDF similarity}
  \label{fig:inter_all}
\end{figure}

In previous experiments, we analyzed the behavior of the system within each trajectory. The results revealed that the multimodality loop does not tend to converge within the same trajectory. Additionally, it was observed that the similarity is lower in the first images compared to the others, thus showing the divergence from the original image. In this experiment, a different approach is studied. Instead of comparing the different image generations within the same trajectory, we compare images from different trajectories but within the same generation (inter-trajectory analysis). One possible result is that all generations have a constant level of similarity since they are independent experiments. Another possibility is that the multimodality loop tends to generate images with certain characteristics (e.g., it transforms all the scenarios with a futuristic style). This would be observed if the similarity increases as the generation number increases. Figure~\ref{fig:inter_all} shows the similarity for the categories as well as for the combination of all the images. It can be observed that the distance remains practically constant across all generations, which indicates that the model does not tend to have common characteristics. The same happens when comparing the descriptions, where the TF-IDF similarity also remains constant. 

\section{Discussion}
\label{Discussion}

The results of the different experiments show that recursive modality changes with state-of-the-art generative AI tools lead to substantial losses in the information, both in image and text. Information loss with modality changes was expected and is consistent with what happens with humans \cite{merino2022textual}. However, in the case of generative AI, the main elements of the initial image are in many cases lost as a result of modality changes even when starting with images that have only a few elements. For example, images of a person degenerate after a few modality change into images that have no persons which is qualitatively different from what would be expected for humans. This occurs for all the image types evaluated containing persons, animals, and different types of objects. There are some elements for which this phenomenon is less likely, for example for persons while for others, like fire hydrants, it occurs frequently. This may be related to the importance of the different objects in the training dataset of the AI tools. In any case, the behavior of generative AI tools is qualitatively different from that of humans.

To understand how information is lost in the recursive modality changes, we have compared the similarity of images and texts across iterations using several metrics. The results show that information is lost faster in the first iterations and then the process continues at a slower pace. In some cases, several iterations maintain some degree of consistency to then depart again to different content. The information loss seems to be larger when changing from text to image than when extracting the text description of an image but the differences are not large. 

One of the limitations of our study is that only a pair of generative AI tools (GPT4o and DALL-E3) has been used. To understand how the tools impact the information loss, additional experiments have been run with different settings of DALL-E3. The results are qualitatively the same, degeneration is also observed. This suggests that it may be a general phenomenon, however additional evaluations with other AI tools are needed to confirm this hypothesis. The same applies to the types of images, more have to be tested to include other objects and also more complex scenes. 

The study presented in this paper is empirical, further work is needed to link the results obtained with the theoretical foundations of the AI models including their architecture and training. This would provide a better understanding of the mechanisms that cause information loss in recursive modality changes and how to improve the AI models to reduce the loss and make the models more stable.

\section{Conclusion and future work}
\label{Conclusion}

In this paper, the recursive changes have been evaluated using the same model in all the iterations. The results show that the multimodal loop tends to diverge from the initial image, with the effect varying more depending on the type of image (category) and the model configuration. This can have a significant impact both on the generation of datasets for model training and on content adaptation, especially as the use of these tools becomes more widespread in society. It would be interesting to study the impact of recursive modality changes when different models are used. For example, different models can be used for each iteration or one model can be used for image generation and a second one for producing text descriptions of the images. Intuitively, the use of different models will make the recursive changes more prone to diverge from the initial content.



%



\section*{Acknowledgments}

This work was supported by the Agencia Estatal de Investigación (AEI) (doi:10.13039/501100011033) under Grant FUN4DATE (PID2022-136684OB-C22), by the European Commission through the Chips Act Joint Undertaking project SMARTY (Grant 101140087), and by the OpenAI Research Program.

\bibliographystyle{IEEEtran}
\bibliography{bare}

\begin{thebibliography}{10}
\providecommand{\url}[1]{#1}
\csname url@samestyle\endcsname
\providecommand{\newblock}{\relax}
\providecommand{\bibinfo}[2]{#2}
\providecommand{\BIBentrySTDinterwordspacing}{\spaceskip=0pt\relax}
\providecommand{\BIBentryALTinterwordstretchfactor}{4}
\providecommand{\BIBentryALTinterwordspacing}{\spaceskip=\fontdimen2\font plus
\BIBentryALTinterwordstretchfactor\fontdimen3\font minus \fontdimen4\font\relax}
\providecommand{\BIBforeignlanguage}[2]{{%
\expandafter\ifx\csname l@#1\endcsname\relax
\typeout{** WARNING: IEEEtran.bst: No hyphenation pattern has been}%
\typeout{** loaded for the language `#1'. Using the pattern for}%
\typeout{** the default language instead.}%
\else
\language=\csname l@#1\endcsname
\fi
#2}}
\providecommand{\BIBdecl}{\relax}
\BIBdecl

\bibitem{ChatGPTOverview}
T.~Wu, S.~He, J.~Liu, S.~Sun, K.~Liu, Q.-L. Han, and Y.~Tang, ``A brief overview of chatgpt: The history, status quo and potential future development,'' \emph{IEEE/CAA Journal of Automatica Sinica}, vol.~10, no.~5, pp. 1122--1136, 2023.

\bibitem{DALLE}
A.~Ramesh, M.~Pavlov, G.~Goh, S.~Gray, C.~Voss, A.~Radford, M.~Chen, and I.~Sutskever, ``Zero-shot text-to-image generation,'' in \emph{International conference on machine learning}.\hskip 1em plus 0.5em minus 0.4em\relax Pmlr, 2021, pp. 8821--8831.

\bibitem{SD1}
R.~Rombach, A.~Blattmann, D.~Lorenz, P.~Esser, and B.~Ommer, ``High-resolution image synthesis with latent diffusion models,'' in \emph{Proceedings of the IEEE/CVF conference on computer vision and pattern recognition}, 2022, pp. 10\,684--10\,695.

\bibitem{inpainting_survey}
W.~Quan, J.~Chen, Y.~Liu, D.-M. Yan, and P.~Wonka, ``Deep learning-based image and video inpainting: A survey,'' \emph{International Journal of Computer Vision}, pp. 1--34, 2024.

\bibitem{MultiModalSurvey}
T.~Baltrušaitis, C.~Ahuja, and L.-P. Morency, ``Multimodal machine learning: A survey and taxonomy,'' \emph{IEEE Transactions on Pattern Analysis and Machine Intelligence}, vol.~41, no.~2, pp. 423--443, 2019.

\bibitem{martinez2023towards}
G.~Mart{\'\i}nez, L.~Watson, P.~Reviriego, J.~A. Hern{\'a}ndez, M.~Juarez, and R.~Sarkar, ``Towards understanding the interplay of generative artificial intelligence and the internet,'' in \emph{International Workshop on Epistemic Uncertainty in Artificial Intelligence}.\hskip 1em plus 0.5em minus 0.4em\relax Springer, 2023, pp. 59--73.

\bibitem{LLMeval}
Z.~Guo, R.~Jin, C.~Liu, Y.~Huang, D.~Shi, L.~Yu, Y.~Liu, J.~Li, B.~Xiong, D.~Xiong \emph{et~al.}, ``Evaluating large language models: A comprehensive survey,'' \emph{arXiv preprint arXiv:2310.19736}, 2023.

\bibitem{kobak2024delving}
D.~Kobak, R.~G. M{\'a}rquez, E.-{\'A}. Horv{\'a}t, and J.~Lause, ``Delving into chatgpt usage in academic writing through excess vocabulary,'' \emph{arXiv preprint arXiv:2406.07016}, 2024.

\bibitem{RAG}
P.~Lewis, E.~Perez, A.~Piktus, F.~Petroni, V.~Karpukhin, N.~Goyal, H.~K\"{u}ttler, M.~Lewis, W.-t. Yih, T.~Rockt\"{a}schel, S.~Riedel, and D.~Kiela, ``Retrieval-augmented generation for knowledge-intensive nlp tasks,'' in \emph{Proceedings of the 34th International Conference on Neural Information Processing Systems}, ser. NIPS '20.\hskip 1em plus 0.5em minus 0.4em\relax Red Hook, NY, USA: Curran Associates Inc., 2020.

\bibitem{Collapse1}
E.~Dohmatob, Y.~Feng, P.~Yang, F.~Charton, and J.~Kempe, ``A tale of tails: Model collapse as a change of scaling laws,'' \emph{arXiv preprint arXiv:2402.07043}, 2024.

\bibitem{Collapse2}
Q.~Bertrand, A.~J. Bose, A.~Duplessis, M.~Jiralerspong, and G.~Gidel, ``On the stability of iterative retraining of generative models on their own data,'' \emph{arXiv preprint arXiv:2310.00429}, 2023.

\bibitem{Collapse3}
M.~Briesch, D.~Sobania, and F.~Rothlauf, ``Large language models suffer from their own output: An analysis of the self-consuming training loop,'' \emph{arXiv preprint arXiv:2311.16822}, 2023.

\bibitem{Collapse4}
M.~Marchi, S.~Soatto, P.~Chaudhari, and P.~Tabuada, ``Heat death of generative models in closed-loop learning,'' \emph{arXiv preprint arXiv:2404.02325}, 2024.

\bibitem{conde2024open}
J.~Conde, M.~Gonz{\'a}lez, N.~Melero, R.~Ferrando, G.~Mart{\'\i}nez, E.~Merino-G{\'o}mez, J.~A. Hern{\'a}ndez, and P.~Reviriego, ``Open source conversational llms do not know most spanish words,'' \emph{arXiv preprint arXiv:2403.15491}, 2024.

\bibitem{conde2024stable}
J.~Conde, M.~Gonz{\'a}lez, G.~Mart{\'\i}nez, F.~Moral, E.~Merino-G{\'o}mez, and P.~Reviriego, ``How stable is stable diffusion under recursive inpainting (rip)?'' \emph{arXiv preprint arXiv:2407.09549}, 2024.

\bibitem{MLLM_survey}
J.~Wu, W.~Gan, Z.~Chen, S.~Wan, and P.~S. Yu, ``Multimodal large language models: A survey,'' in \emph{2023 IEEE International Conference on Big Data (BigData)}, 2023, pp. 2247--2256.

\bibitem{MLLM_survey2}
S.~Yin, C.~Fu, S.~Zhao, K.~Li, X.~Sun, T.~Xu, and E.~Chen, ``A survey on multimodal large language models,'' \emph{arXiv preprint arXiv:2306.13549}, 2023.

\bibitem{gpt4}
J.~Achiam, S.~Adler, S.~Agarwal, L.~Ahmad, I.~Akkaya, F.~L. Aleman, D.~Almeida, J.~Altenschmidt, S.~Altman, S.~Anadkat \emph{et~al.}, ``Gpt-4 technical report,'' \emph{arXiv preprint arXiv:2303.08774}, 2023.

\bibitem{team2023gemini}
G.~Team, R.~Anil, S.~Borgeaud, Y.~Wu, J.-B. Alayrac, J.~Yu, R.~Soricut, J.~Schalkwyk, A.~M. Dai, A.~Hauth \emph{et~al.}, ``Gemini: a family of highly capable multimodal models,'' \emph{arXiv preprint arXiv:2312.11805}, 2023.

\bibitem{bai2023qwen}
J.~Bai, S.~Bai, S.~Yang, S.~Wang, S.~Tan, P.~Wang, J.~Lin, C.~Zhou, and J.~Zhou, ``Qwen-vl: A versatile vision-language model for understanding, localization, text reading, and beyond,'' 2023.

\bibitem{InternVL}
Z.~Chen, W.~Wang, H.~Tian, S.~Ye, Z.~Gao, E.~Cui, W.~Tong, K.~Hu, J.~Luo, Z.~Ma \emph{et~al.}, ``How far are we to gpt-4v? closing the gap to commercial multimodal models with open-source suites,'' \emph{arXiv preprint arXiv:2404.16821}, 2024.

\bibitem{LLava}
H.~Liu, C.~Li, Q.~Wu, and Y.~J. Lee, ``Visual instruction tuning,'' \emph{Advances in neural information processing systems}, vol.~36, 2024.

\bibitem{dubey2024llama}
A.~Dubey, A.~Jauhri, A.~Pandey, A.~Kadian, A.~Al-Dahle, A.~Letman, A.~Mathur, A.~Schelten, A.~Yang, A.~Fan \emph{et~al.}, ``The llama 3 herd of models,'' \emph{arXiv preprint arXiv:2407.21783}, 2024.

\bibitem{MedievalAnimals1}
A.~Pluskowski, ``Narwhals or unicorns? exotic animals as material culture in medieval europe,'' \emph{European Journal of Archaeology}, vol.~7, pp. 291--313, 12 2004.

\bibitem{Bestiary}
L.~Grollemond, \emph{Book of Beasts: The Bestiary in the Medieval World}.\hskip 1em plus 0.5em minus 0.4em\relax Getty Publications, 2019.

\bibitem{ogasawara2021durer}
L.~Ogasawara, ``D{\"u}rer's rhinoceros,'' \emph{Pleiades: Literature in Context}, vol.~41, no.~2, pp. 140--144, 2021.

\bibitem{birkin2024durer}
J.~Birkin and S.~Manghani, ``From d{\"u}rer’s rhinoceros to ai image diffusion models,'' \emph{Holotipus rivista di zoologia sistematica e tassonomia}, vol.~5, no.~1, pp. 3--19, 2024.

\bibitem{Babel}
F.~Condorelli, B.~Tramelli, and A.~Luigini, ``Digital turris babel. augmented release of athanasius kircher’s archontologia,'' in \emph{Beyond Digital Representation: Advanced Experiences in AR and AI for Cultural Heritage and Innovative Design}.\hskip 1em plus 0.5em minus 0.4em\relax Springer, 2023, pp. 83--95.

\bibitem{Solomon}
J.~M. Lundquist, \emph{The Temple of Jerusalem: past, present, and future}.\hskip 1em plus 0.5em minus 0.4em\relax Bloomsbury Publishing USA, 2007.

\bibitem{Crete}
V.-P. Herva and J.~Rapakko, ``Insides, outsides and the labyrinth: Knossos, palatial space and environmental perception in minoan crete,'' \emph{Journal of Social Archaeology}, vol.~23, no.~3, pp. 264--285, 2023.

\bibitem{merino2024word}
E.~Merino-G{\'o}mez, F.~Moral-Andr{\'e}s, B.~Querol, P.~Reviriego, and S.~Pisaniello, ``Word pictures: New insights through ai around the villa laurentina by pliny the younger,'' \emph{Decoding Cultural Heritage: A Critical Dissection and Taxonomy of Human Creativity through Digital Tools}, pp. 47--68, 2024.

\bibitem{coco_dataset}
T.-Y. Lin, M.~Maire, S.~Belongie, J.~Hays, P.~Perona, D.~Ramanan, P.~Doll{\'a}r, and C.~L. Zitnick, ``Microsoft coco: Common objects in context,'' in \emph{Computer Vision -- ECCV 2014}, D.~Fleet, T.~Pajdla, B.~Schiele, and T.~Tuytelaars, Eds.\hskip 1em plus 0.5em minus 0.4em\relax Cham: Springer International Publishing, 2014, pp. 740--755.

\bibitem{LPIPS}
R.~Zhang, P.~Isola, A.~A. Efros, E.~Shechtman, and O.~Wang, ``The unreasonable effectiveness of deep features as a perceptual metric,'' in \emph{Proceedings of the IEEE conference on computer vision and pattern recognition}, 2018, pp. 586--595.

\bibitem{blip}
J.~Li, D.~Li, C.~Xiong, and S.~Hoi, ``{BLIP}: Bootstrapping language-image pre-training for unified vision-language understanding and generation,'' in \emph{Proceedings of the 39th International Conference on Machine Learning}, ser. Proceedings of Machine Learning Research, K.~Chaudhuri, S.~Jegelka, L.~Song, C.~Szepesvari, G.~Niu, and S.~Sabato, Eds., vol. 162.\hskip 1em plus 0.5em minus 0.4em\relax PMLR, 17--23 Jul 2022, pp. 12\,888--12\,900.

\bibitem{TF-IDF}
K.~Sparck~Jones, ``A statistical interpretation of term specificity and its application in retrieval,'' \emph{Journal of documentation}, vol.~28, no.~1, pp. 11--21, 1972.

\bibitem{xin2024dart}
C.~Xin, A.~Hartel, and E.~Kasneci, ``Dart: An automated end-to-end object detection pipeline with data diversification, open-vocabulary bounding box annotation, pseudo-label review, and model training,'' \emph{arXiv preprint arXiv:2407.09174}, 2024.

\bibitem{merino2022textual}
E.~Merino~Gomez and F.~Moral~Andres, ``Textual architectures: Visual invention through narrative reception,'' \emph{RA-REVISTA DE ARQUITECTURA}, no.~24, pp. 259--265, 2022.

\end{thebibliography}

\appendix

\clearpage
\section{Annex: Summary of multimodal loops}

\begin{figure}[h]
  \centering
  \includegraphics[scale=0.20]{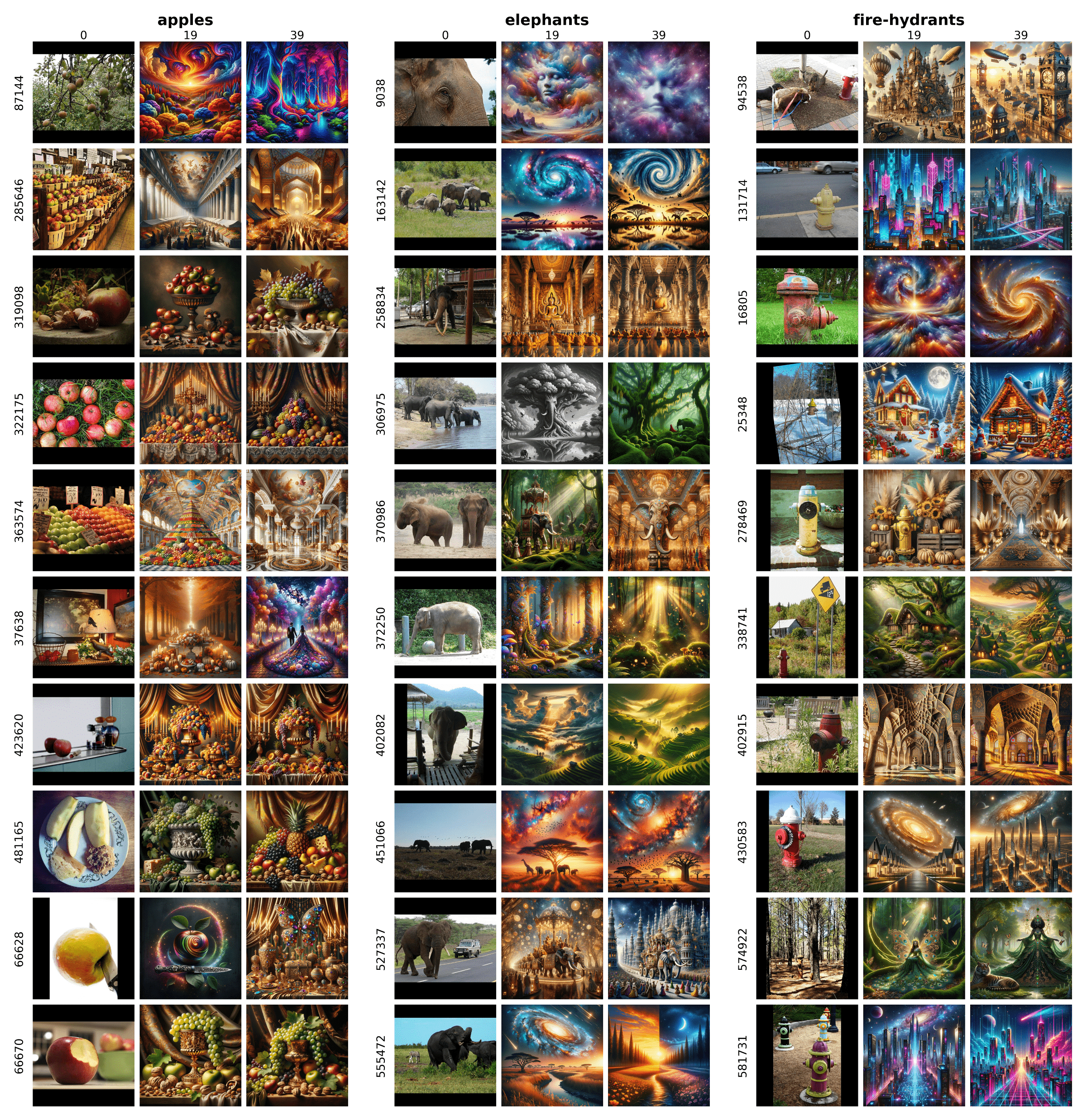}
  \caption{Examples of all the trajectories (categories apples, elephants, and fire-hydrants). Every image includes the generations 0, 19, and 39.}
  \label{fig:grids_summary_1}
\end{figure}

\begin{figure}[h]
  \centering
  \includegraphics[scale=0.20]{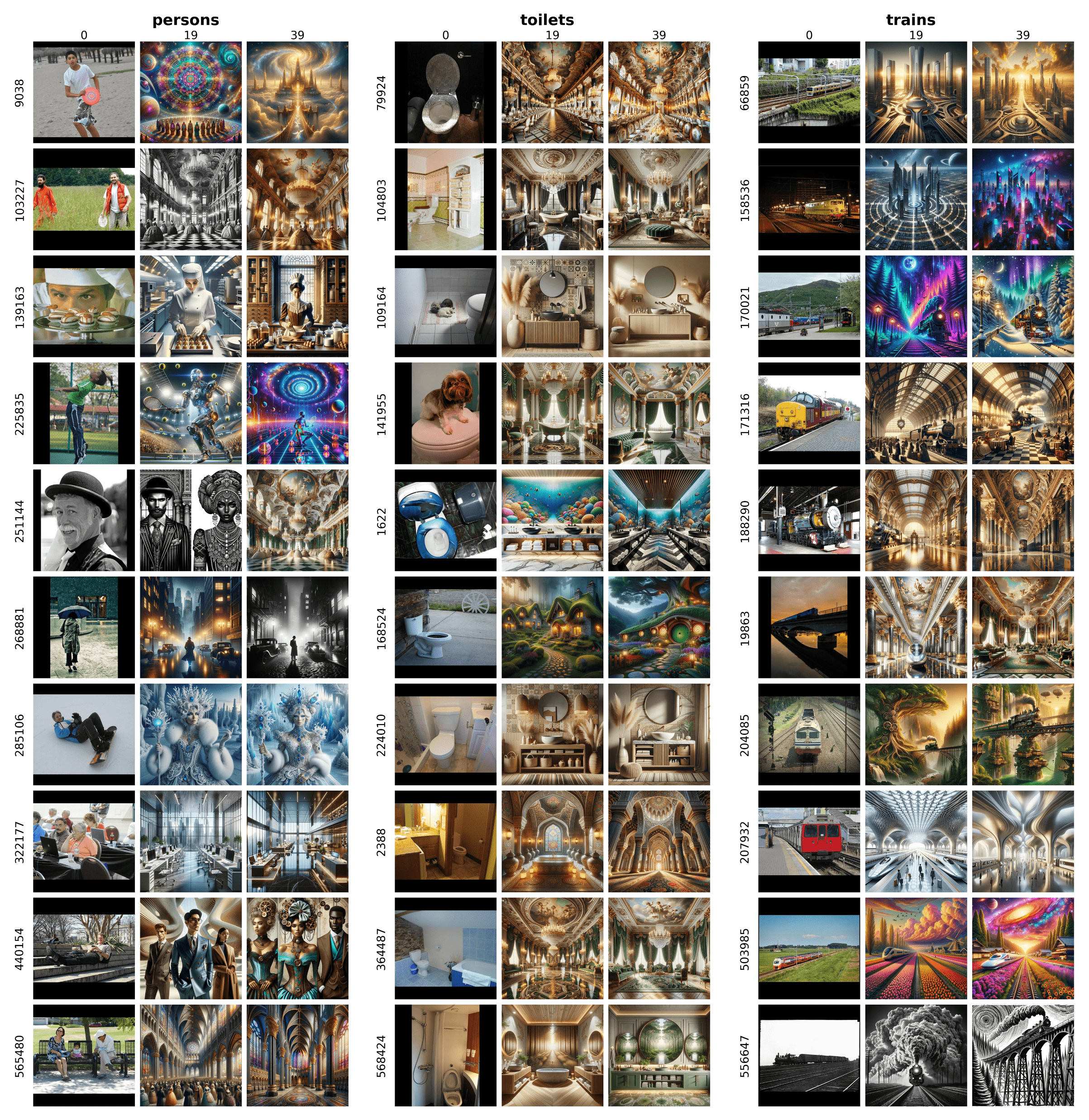}
  \caption{Examples of all the trajectories (categories: persons, toilets, and trains). Every image includes the generations 0, 19, and 39.}
  \label{fig:grids_summary_2}
\end{figure}

\end{document}